\documentclass[
  journal=largetwo,
  manuscript=article-type,
  year=2024,
  volume=37,
]{cup-journal}

\usepackage{amsmath}
\usepackage{amsfonts}
\usepackage{amssymb}
\usepackage[nopatch]{microtype}
\usepackage{booktabs}
\usepackage[table]{xcolor}
\usepackage{multirow}
\usepackage{aas_macros}

\defcitealias{NFW1995}{NFW 1995}
\defcitealias{NFW1996}{NFW 1996}
\defcitealias{NFW1997}{NFW 1997}

\usepackage{soul}

\title{Predicting the Scaling Relations between the Dark Matter Halo Mass and Observables from Generalised Profiles II: Intracluster Gas Emission}

\author{A. Sullivan}
\affiliation{International Centre for Radio Astronomy Research, The University of Western Australia, 35 Stirling Highway, Crawley, Western Australia, 6009, Australia}
\alsoaffiliation{ARC Centre of Excellence for All Sky Astrophysics in 3 Dimensions (ASTRO 3D)}

\author{C. Power}
\affiliation{International Centre for Radio Astronomy Research, The University of Western Australia, 35 Stirling Highway, Crawley, Western Australia, 6009, Australia}
\alsoaffiliation{ARC Centre of Excellence for All Sky Astrophysics in 3 Dimensions (ASTRO 3D)}

\author{C. Bottrell}
\affiliation{International Centre for Radio Astronomy Research, The University of Western Australia, 35 Stirling Highway, Crawley, Western Australia, 6009, Australia}

\author{A. Robotham}
\affiliation{International Centre for Radio Astronomy Research, The University of Western Australia, 35 Stirling Highway, Crawley, Western Australia, 6009, Australia}
\alsoaffiliation{ARC Centre of Excellence for All Sky Astrophysics in 3 Dimensions (ASTRO 3D)}

\author{S. Shabala}
\affiliation{School of Natural Sciences, University of Tasmania, Hobart, Tasmania, Australia}
\email[A. Sullivan]{andrew.sullivan@icrar.org}


\begin{document}

\begin{abstract}
    We investigate the connection between a cluster's structural configuration and observable measures of its gas emission that can be obtained in X-ray and Sunyaev-Zeldovich (SZ) surveys. We present an analytic model for the intracluster gas density profile: parameterised by the dark matter halo’s inner logarithmic density slope, $\alpha$, the concentration, $c$, the gas profile's inner logarithmic density slope, $\varepsilon$, the dilution, $d$, and the gas fraction, $\eta$, normalised to cosmological content. We predict four probes of the gas emission: the emission-weighted, $T_\mathrm{X}$, and mean gas mass-weighted, $T_\mathrm{m_g}$, temperatures, and the spherically, $Y_\mathrm{sph}$, and cylindrically, $Y_\mathrm{cyl}$, integrated Compton parameters. Over a parameter space of clusters, we constrain the X-ray temperature scaling relations, $M_{200} - T_\mathrm{X}$ and $M_{500} - T_\mathrm{X}$, within $57.3\%$ and $41.6\%$, and $M_{200} - T_\mathrm{m_g}$ and $M_{500} - T_\mathrm{m_g}$, within $25.7\%$ and $7.0\%$, all respectively. When excising the cluster's core, the $M_{200} - T_\mathrm{X}$ and $M_{500} - T_\mathrm{X}$ relations are further constrained, to within $31.3\%$ and $17.1\%$, respectively. Similarly, we constrain the SZ scaling relations, $M_{200} - Y_\mathrm{sph}$ and $M_{500} - Y_\mathrm{sph}$, within $31.1\%$ and $17.7\%$, and $M_{200} - Y_\mathrm{cyl}$ and $M_{500} - Y_\mathrm{cyl}$, within $25.2\%$ and $22.0\%$, all respectively. The temperature observable $T_\mathrm{m_g}$ places the strongest constraint on the halo mass, whilst $T_\mathrm{X}$ is more sensitive to the parameter space. The SZ constraints are sensitive to the gas fraction, whilst insensitive to the form of the gas profile itself. In all cases, the halo mass is recovered with an uncertainty that suggests the cluster's structural profiles only contribute a minor uncertainty in its scaling relations. 
\end{abstract}

\section{1. \quad Introduction}

Our currently favoured standard cosmological model predicts that dark matter halos are the building blocks of structure formation and host the galaxies, galaxy groups, and clusters that we observe \autocite[e.g.][]{White1978, FrenkWhite1991}. At the mass scale of galaxy clusters, the halo's mass can be observationally inferred from the thermodynamic properties of its hot gaseous atmosphere, either from its X-ray emission \autocite[e.g.][]{Vikhlinin2006, Vikhlinin2009, Babyk2023} or from the distortion of the Cosmic Microwave Background (CMB) due to interactions of CMB photons with energetic electrons \autocite[e.g.][]{Vanderlinde2010, Andersson2011}, known as the Sunyaev-Zeldovich \autocite[SZ;][]{SZ1970, SZ1972} effect. 



Observationally, the advent of precision X-ray telescopes such as XMM-\textit{Newton} \autocite[e.g.][]{XMMNewton}, \textit{Chandra} \autocite[e.g.][]{Chandra}, and eROSITA \autocite[e.g.][]{eROSITA} have allowed for halo mass estimates to be routinely obtained from statistical samples of X-ray emitting galaxy clusters. These X-ray derived halo mass estimates typically rely on multi-parameter 3-dimensional fits to the gas density and temperature profiles, from which the total halo mass within a halocentric radius is estimated by requiring hydrostatic equilibrium \autocite[i.e. pressure forces balance gravitational forces; see, e.g.][]{Sarazin1988}. 
These halo masses can be correlated with mean-weighted temperature observables, typically weighted by either the X-ray emission or the gas mass, to establish scaling relations with the halo mass \autocite[see, e.g.][]{Arnaud2005, Vikhlinin2006, Vikhlinin2009, Babyk2023}.

Similarly, SZ-selected cluster observations have been catalogued by precision microwave telescopes such as the South Pole Telescope (SPT) \autocite[e.g.][]{SPT}, the Atacama Cosmology Telescope (ACT) \autocite[e.g.][]{ACT} and \textit{Planck} \autocite[e.g.][]{Planck2015}. These surveys allow the integrated SZ signal, also called the Compton parameter, to be measured for galaxy clusters, which when combined with X-ray cluster observations permitting hydrostatic mass estimates, can establish the scaling relation between the cluster's SZ signal and its halo mass \autocite[e.g.][]{Andersson2011, Czakon2015, Liu2015}. 

Generally, fits to a cluster's X-ray emission allow tight limits to be placed on the halo mass, up to a hydrostatic bias, which is thought to underestimate the halo mass in relaxed galaxy clusters by $\sim 10-20\%$ \autocite[e.g.][]{Martizzi2016, Ettori2022}. However, it is notable that these hydrostatic mass estimates rely on fitting multi-parameter models to the gas emission, which are often empirically rather than physically motivated, producing a variety of `universal forms' for the gas density profile \autocite[e.g.][]{Vikhlinin2006, Pratt2023, Lyskova2023}, the gas temperature profile \autocite[e.g.][]{Sun2009, Ghirardini2019} and the gas pressure profile \autocite[e.g.][]{Arnaud2010}. This reflects both the variance in the forms of gas profiles devised from observational data, and the complexity of the underlying physical processes \autocite[e.g. cosmological gas accretion, infalling groups, outflows of feedback; see, e.g.][]{Power2020} that shape these profiles. Moreover, these empirical fits are often at odds with theoretically motivated, analytic models for the gas profile \autocite[e.g.][]{Cavaliere&Fusco-Femiano1978, Komatsu&Seljak2001} which struggle to capture empirical results.

\textcolor{black}{
Simple parametric models of galaxy clusters have been constructed to predict their emission properties, as informed by empirical and numerical results \autocite[e.g.][]{Bode2009, Allison2011}. 
Similar work has investigated the nature of galaxy cluster scaling relations; in particular, applying Bayesian approaches to constraining these relations from observational data \autocite[e.g.][]{Maughan2014}, and developing semi-analytic models that explain observed deviations of these scaling relations from self-similarity \autocite[e.g.][]{Ettori2015, Ettori2023}. More recently, algorithms for `baryon pasting' gas profiles onto numerically simulated dark matter-only halos have generated mock observations, particularly for SZ observables \autocite[e.g.][]{Osato2023}, exciting the prospect of constraining the halo mass, when informed by its reconstruction from these mock observables.}


\textcolor{black}{
In these parametric and statistical approaches to studying galaxy clusters and their scaling relations, the structures of the cluster's dark matter and intracluster gas components are assumed, and their relation to the hydrostatic state of the system. In this work, we investigate how these scaling relations depend on cluster structure and composition, and how much they are expected to vary, when the cluster is parameterised by relatively agnostic prescriptions for its structure. This study builds on our previous work (Sullivan et al. 2024, Submitted), where we investigated the scaling relation between the kinematic observables of tracer populations within a halo and the underlying halo mass. In this work, we turn to the scaling of X-ray and SZ observables with the halo mass, specifically at the mass scale of galaxy clusters, and within the regime of self-similarity. To achieve this, one of our key goals is to construct a physically motivated, simple analytical profile to model galaxy clusters and their intracluster gas component and to accurately predict their emission observables in terms of the cluster's structural parameters.}


Our analytic tool-kit is detailed in Section 2: we review theoretical predictions and empirically motivated forms for the density profiles of the intracluster gas and the dark matter within galaxy clusters, and postulate a generalised profile which we call the `ideal baryonic cluster halo'. Thereafter, we use the virial theorem to constrain the correspondence between a cluster's halo mass and its emission observables: namely, its mean-weighted temperatures and integrated SZ signals. In Section 3 the analytic profiles for these observables are derived, with an analysis of their bounds over the outlined parameter space fixing constraints on the halo mass. Scaling relations are presented in Section 4, along with a discussion on the dependence of our results on the chosen parameters built into the cluster's structural model. We present our conclusions in Section 5.

\section{2. \quad Theoretical background and methods}
\vspace{2mm}
\subsection{2.1. \quad Constructing a generalised dark matter profile}

In the first paper of our series (Sullivan et al. 2023, referred to as Paper I hereafter), we modelled halos as dark matter-only systems, using a generalisation of the Navarro-Frenk-White profile (\cite{NFW1995, NFW1996, NFW1997}, hereafter NFW) which has been found to provide a good fit to the ensemble average of dynamically relaxed halos in cosmological $N$-body simulations. 
We referred to this generalisation as `ideal physical halos'; they describe spherically symmetric halos with logarithmic density slope (hereafter, for brevity, slope) of $-3$ at large radius, and inner slope $-\alpha$, such that the density profile is parameterised by $\rho(r) \sim r^{-\alpha}$ at small radii. This form was motivated as a reasonable approximation to relaxed and unperturbed halos in both the observed and simulated universe. 

\subsubsection{The dark matter halo inner slope $\alpha$}

The NFW profile is the $\alpha=1$ member of the ideal physical halo class, with its divergent behaviour at small radii referred to as a `cusp'. If the density profile were instead to flatten in the inner region, as $\alpha \simeq 0$, this is referred to as a `core'. NFW-like cusps are consistently predicted in cosmological $N$-body simulations, while even `cuspier' inner slopes of $\alpha \simeq 1.5$ have been recovered in recent high-resolution idealised $N$-body simulations that follow the collapse of proto-halos \autocite[e.g.][]{OgiyaHahn2017}. However, this is in tension with observational evidence of dark matter dominated galaxies,  such as dwarfs and low-surface brightness galaxies \autocite[e.g.][]{Moore1994,deblok1997,deblok2001,kuziodenaray2008}{}{}, which favour central cores; this is often referred to as the `core-cusp problem'. We note that there is evidence from cosmological hydrodynamical simulations that model galaxy formation processes (e.g. gas cooling, stellar feedback) that these can disrupt cusps \autocite[][]{oh2011,pontzen2012,dicintio2014a,dicintio2014b}{}{} and produce cored central densities. For these reasons, in Paper I we chose to model the ideal physical halos within the parameter space of halo inner slopes $\alpha \in [0, 1.5]$, encompassing a plausible range of predictions. 

\subsubsection{The concentration parameter $c$}

The properties of dark matter halos are usually referenced in terms of their so-called virial parameters. These parameters describe gravitational structures in virial equilibrium, the state in which its gravitational potential energy is balanced by its internal energy (approximately; cf. \cite{colelacey1996}), as expected to be established within halos, predicted from gravitational collapse models. The virial mass, $M_\mathrm{vir}$, defines the halo mass enclosed within a sphere of virial radius, $r_\mathrm{vir}$, as:
\begin{equation}\label{virial mass definition}
    M_\mathrm{vir} \equiv \frac{4}{3} \pi r_\mathrm{vir}^3 \Delta \rho_\mathrm{crit,0},
\end{equation}
such that the mean enclosed mass density is given by the overdensity parameter, $\Delta$, times the present critical density of the universe, $\rho_\mathrm{crit,0}$ \autocite[e.g.][]{white2001}{}{}. 

A standard convention in the literature is that $\Delta=200$, defining
$M_{200}$ and 
$r_{200}$ to approximate a halo's virial mass and radius, respectively. In X-ray observations, where temperature measurements are usually limited to the higher density, smaller radii, regions of a halo's gas distribution, it is often convenient 
to choose $\Delta=500$, corresponding to the halo mass $M_{500}$ and the halo radius $r_{500}$.
In our investigation, we will estimate the relationship between the observable properties of clusters and the halo mass, in both of the conventions $M_{200}$ and $M_{500}$. 

When normalising a dark matter halo's density profile to contain $M_\mathrm{vir}$ within $r_\mathrm{vir}$, the concentration parameter, $c$, is typically introduced to parameterise the profile, defined as the ratio of the outer virial radius, $r_\mathrm{vir}$, to the inner scale radius, $r_\mathrm{s}$, as:
\begin{equation}\label{concentration definition}
    c \equiv \frac{r_\mathrm{vir}}{r_\mathrm{s}}.
\end{equation}
The concentration parameter is found in cosmological $N$-body simulations to be weakly dependent on halo mass, with $c=5$ corresponding to cluster-scale halos in the standard $\Lambda \mathrm{CDM}$ cosmology \autocite[e.g.][]{bullock2001,duffy2008,ludlow2012,ludlow2014}{}{}. 

The definition of the concentration parameter $c$ depends on the choice in overdensity, $\Delta$. As $r_{500} < r_{200}$, by definition in Equation \eqref{concentration definition}, this implies an overdensity-dependent concentration, whereby $c_{500} < c_{200}$. In this paper, we refer to the concentration only as $c$, and take suitable values when taking different choices of $\Delta$. In particular, we take the typical $\Lambda \mathrm{CDM}$ cluster-scale concentration $c=5$ when $\Delta=200$, and assume to first order $r_{500}/r_{200} \simeq 0.5$\footnote{In general, the value of $r_{500}/r_{200}$ for a given halo will depend on the specific form of its density profile. This exact conversion is not of strong importance in our model, as $r_{200}$ and $r_{500}$ are both assumed to be virial radius approximations, and so any modification of their ratio will be quantitative rather than qualitative.}, so that $c=2.5$ appropriately describes cluster-scale halos when $\Delta=500$. 

\subsubsection{The ideal physical halo profile}
To model a halo's density profile in a scale-free formalism, a dimensionless radial scale, $s$, can be introduced, defined as:
\begin{equation}\label{dimensionless radius definition}
    s \equiv \frac{r}{r_\mathrm{vir}},
\end{equation}
where $r$ is the halocentric radius. As per Paper I, this allowed us to express the density profile of the ideal physical halos in the form:

\begin{equation}\label{ideal physical halo profile}
    \frac{\rho (s, c, \alpha)}{\Delta \rho_\mathrm{crit, 0}} = \frac{u(c, \alpha)}{3s^\alpha (1 + cs)^{3-\alpha}},
\end{equation}
where $c$ is the concentration, $\alpha$ is the halo's inner slope, and we refer to $u(c, \alpha)$ as the generalised concentration function, defined by the integral:
\begin{equation}\label{ideal physical halo concentration function}
    u(c, \alpha) \equiv \left[\int _0 ^1 \frac{s^{2-\alpha} \mathrm{d}s}{(1 + cs)^{3-\alpha}}\right]^{-1}.
\end{equation}

\subsection{2.2. \quad Constructing a generalised intracluster gas profile}

We begin this work by constructing 
a gas density profile, $\rho_\mathrm{gas}(r)$, to model the distribution of hot, X-ray emitting ionised gas within a galaxy cluster.
For now, we assume that the halo's underlying dark matter density profile, $\rho_\mathrm{dm}(r)$, takes the form given by the ideal physical halo profile in Equation \eqref{ideal physical halo profile}, parameterised by the halo's concentration and inner slope. 

To model the gas distribution within a cluster, historically the isothermal-$\beta$ profile has been the preferred model: taken to model a spherically symmetric isothermal gas profile with a central core \autocite{Cavaliere&Fusco-Femiano1978}. However, X-ray surveys have demonstrated that the temperature distribution within clusters is far from isothermal, when parameterised as function of halocentric radius \autocite[e.g.][]{Vikhlinin2006, Sun2009, Lyskova2023}. These X-ray studies instead devise a cluster-averaged intracluster gas density profile, when combining radially parameterised fits for each cluster within a given survey; typically, this recovers a gas profile described by dual inner and outer logarithmic density slopes, with general consensus for the outer slope nearing $-3$ toward limiting radii \autocite[e.g.][]{Vikhlinin2006, Croston2008, Lyskova2023, Pratt2023}. 

In this study, we parameterise a scale-free gas density profile in terms of a minimal set of physically-understood parameters. To do this, we will assume an ansatz for this profile, of the form:
\begin{equation}\label{Ansatz}
    \frac{\rho_\mathrm{gas}(s, \varepsilon)}{\Delta \rho_\mathrm{crit,0}} = \frac{\delta_\mathrm{char}}{3s^\varepsilon (1 + \mathcal{C}s)^{3-\varepsilon}};
\end{equation}
or, equivalently, that the gas density is encompassed by the set of all spherically symmetric gas profiles with an outer slope of $-3$, and an inner slope of $-\varepsilon$, such that $\rho_\mathrm{gas}(r) \sim r^{-\varepsilon}$ models the gas distribution in the inner region of the cluster. In this assumed form, there are two additional terms: the profile's characteristic density, $\delta_\mathrm{char}$, and the parameter $\mathcal{C}$, which each represent undetermined normalisation factors. From this ansatz, we will seek to constrain expressions for $\delta_\mathrm{char}$ and $\mathcal{C}$ in terms of structural parameters for the intracluster gas, such that the form of this density profile is well-understood and physically grounded. 

\subsubsection{The dilution parameter $d$}

The defining assumption built into the ansatz in Equation \eqref{Ansatz} is that the logarithmic slope of the gas and the dark matter profile converge at some outer radius. This must occur, as the characteristic form imposed for these profiles imposes that each attain an outer slope of $-3$ beyond some radial scale. This assumption is justified from modelling provided by hydrodynamic simulations, whereby this convergence is unanimously recovered in galaxy clusters \autocite[see, e.g.][]{Frenk1999}.

The radius in which these density slopes first converge we term the dilution radius, $r_\mathrm{dil}$, describing the point where the gas slope is `diluted' to the same slope as the underlying halo. To encode this structural property, we define the dilution parameter, $d$, as the ratio of the virial radius, $r_\mathrm{vir}$ to this dilution radius, $r_\mathrm{dil}$, as:
\begin{equation}\label{dilution parameter definition}
    d \equiv \frac{r_\mathrm{vir}}{r_\mathrm{dil}}.
\end{equation}
From hydrodynamic simulations of $\Lambda \mathrm{CDM}$ cluster halos, this dilution of the gas typically occurs at or near $r_\mathrm{dil} \simeq 0.5 r_\mathrm{200}$ \autocite[e.g.][]{Yoshikawa2000}. Consequently, when taking an overdensity $\Delta=200$, we will assume a dilution of $d=2$ to model the intracluster gas profile. Importantly, just as the halo concentration, $c$, was overdensity-dependent, by its definition in Equation \eqref{dilution parameter definition}, the dilution, $d$, is also overdensity-dependent. In the same way, we will assume a factor of $0.5$ when taking an overdensity $\Delta=500$, such that the corresponding dilution is $d=1$ in this instance. 

With this definition of the dilution parameter, we can take the logarithmic derivatives of the expected dark matter halo profile from Equation \eqref{ideal physical halo profile} and the assumed form of the gas profile from Equation \eqref{Ansatz}, and impose equality for $r\geq r_\mathrm{dil}$, or in their scale-free composition, for $s \geq 1/d$. By imposing this condition, the parameter $\mathcal{C}$ can be constrained as:
\begin{equation}\label{baryon concentration parameter}
    \mathcal{C}(c, \alpha, d, \varepsilon) \equiv \frac{d(\alpha - \varepsilon) + c(3 - \varepsilon)}{3-\alpha},
\end{equation}
as specified by the parameters $c$ and $\alpha$ that describe the dark matter halo profile, and the dilution, $d$. We will refer to this function $\mathcal{C}(c, \alpha, d, \varepsilon)$ as the gas' concentration parameter, as it appears in analogue to the halo's concentration parameter, $c$, in the form of the density profiles.

\subsubsection{The fraction of cosmological baryon content $\eta$}

When normalising the density profiles of a cluster comprised of both dark matter and intracluster gas, we must ensure that the virial mass is defined by integrating the total density ($\rho_\mathrm{dm} + \rho_\mathrm{gas}$) out to the virial radius, rather than only the dark matter component, as was done in Paper I. To do this, we must introduce a measure of the relative contribution of gas to dark matter in contributing to the halo's total mass; in this case, we will parameterise the fraction of gas mass to total mass, $f_\mathrm{gas}(r \leq r_\mathrm{vir})$, as measured at the cluster's virial radius, $r_\mathrm{vir}$. 

The present-day cosmological value of the fraction of baryonic mass (stars  + gas) to total mass in the universe is called the cosmological baryon fraction, $f_\mathrm{b, cos}$, defined in the ratio:
\begin{equation}\label{cosmological baryon fraction}
    f_\mathrm{b, cos} \equiv \frac{\Omega_\mathrm{b,0}}{\Omega_\mathrm{m,0}},
\end{equation}
of the present-day baryonic density parameter, $\Omega_\mathrm{b,0}$, to the present-day total mass density parameter, $\Omega_\mathrm{m,0}$. This cosmological parameter has been constrained from precision measurement of anisotropies on the Cosmic Microwave Background. We take the value $f_\mathrm{b, cos} = 0.158$ \autocite[]{Planck2016} whilst neglecting uncertainties, as these will be insignificant compared to the range spanned by the parameter space of the gas profiles. 

Although our cluster halo model does not consider stars, it is thought that the hot intracluster gas component dominates the baryonic composition of X-ray emitting clusters \autocite[see, e.g.][]{Akino2022}. This allows us to apply a cosmological normalisation to our gas profiles, by defining the parameter $\eta$ as the ratio of the halo cluster's gas fraction to the cosmological baryon fraction, as:
\begin{equation}\label{fraction of cosmological baryon content}
    \eta \equiv \frac{f_\mathrm{gas}(r \leq r_\mathrm{vir})}{f_\mathrm{b, cos}},
\end{equation}
where $r_\mathrm{vir}$ is taken as either $r_{200}$ or $r_{500}$, depending on the choice in overdensity, $\Delta$. We refer to this parameter, $\eta$, as the halo's fraction of cosmological baryon content. 

Observational measurements for the gas fraction of clusters is routinely devised in X-ray observations, from fits to the gas density profile, where this fraction is consistently found to be an increasing function with cluster halo mass \autocite[e.g.][]{Arnaud2005, Vikhlinin2009}. Whilst early X-ray surveys measured the mean gas fraction for cluster samples around $f_\mathrm{gas}(\leq r_{500}) \simeq 0.106-0.110$, corresponding to $\eta \simeq 0.67-0.70$ \autocite[e.g.][]{Vikhlinin2006, Ettori2009}, recent, more precisely calibrated surveys have consistently measured a higher mean gas fraction of $f_\mathrm{gas} (\leq r_{500}) \simeq 0.130-0.132$, corresponding to $\eta \simeq 0.82-0.84$ \autocite[e.g.][]{Eckert2013, Morandi2015, Pratt2023}. At the cluster outskirts, the mean gas fraction is found to be an increasing function of halocentric radius, and its mean value at $r_{200}$ has been estimated around $f_\mathrm{gas}(\leq r_{200}) \simeq 0.150$, close to the cosmological value $\eta \simeq 1$ \autocite[e.g.][]{Morandi2015, Lyskova2023}. 

Encompassing this range of observational predictions --- and remaining agnostic to the choice in measuring the gas fraction within either $r_{200}$ or $r_{500}$, so that this parameter is not overdensity-dependent --- we take the values of $\eta \in [0.6, 1]$ in our parameter space of idealised halos. In terms of this parameterisation, we can normalise the gas content in our cluster halo model by integrating the gas density profile from Equation \eqref{Ansatz} up to $r_\mathrm{vir}$, and setting the enclosed mass to contain a fraction of $\eta$ times the virial mass, $M_\mathrm{vir}$. From this condition, we can constrain the profile's characteristic density, $\delta_\mathrm{char}$, in the form:
\begin{equation}\label{baryon profile characteristic density}
    \delta_\mathrm{char} \equiv \eta f_\mathrm{b, cos} \mathcal{U}(c, \alpha, d, \varepsilon),
\end{equation}
where we refer to $\mathcal{U}(c, \alpha, d, \varepsilon)$ as the gas' generalised concentration function, as defined by the integral expression:
\begin{equation}\label{baryon concentration function}
	\mathcal{U}(c, \alpha, d, \varepsilon) \equiv \left[\int _0 ^{1} \frac{s^{2-\varepsilon}\mathrm{d}s}{[1 + \mathcal{C}(c, \alpha, d, \varepsilon) s]^{3-\varepsilon}} \right] ^{-1},
\end{equation}
with this function appearing in complete analogue to the halo's generalised concentration function, $u(c, \alpha)$, as defined in Equation \eqref{ideal physical halo concentration function}.

\subsubsection{The gas inner slope $\varepsilon$}

The remaining free parameter in Equation \eqref{Ansatz} is $\varepsilon$, the inner slope of the gas profile. Recent fits for the average density profile of the intracluster gas using XMM-\textit{Newton} \autocite{Pratt2023} and eROSITA \autocite{Lyskova2023} have observed weak cusps of $\varepsilon \simeq 0.4-0.7$ in the central regions $r\lesssim 0.1r_{500}$ of clusters, in contrast to the central core modelled in the isothermal-$\beta$ profile and assumed in some analytic models for the gas density \autocite[e.g.][]{Komatsu&Seljak2001}. More comprehensive studies have observed a large scatter in the behaviour of the gas profile in this central region \autocite[see, e.g.][]{Croston2008}, with no clear trend between this inner structure of the gas and the cluster's macroscopic properties. 

Numerical studies have suggested that the behaviour of gas profiles in the central region of clusters are dominated by non-gravitational feedback processes \autocite[e.g.][]{McDonald2017}. Analytic modelling has been employed to attempt to relate this inner gas structure to the gas accretion properties at or near the virial radius: in particular, by modelling the shock-wave propagated across this boundary. This approach has established mathematical correlations between the gas inner slope and the strength of this shock \autocite[see, e.g.][]{Patej&Loeb2015}; however, it is not clear if this behaviour in real halos is instead dominated by more complex, non-analytic feedback processes. 

In this study, we will model the intracluster gas density profile by inner slopes within the range $\varepsilon \in [0, 1]$: encompassing the predictions from recent observational fits, whilst permitting a generous scatter in values, as anticipated for real clusters undergoing complex processes in the central regions. 

\begin{figure}[h!]
    \centering
    \caption{The intracluster gas density profiles, in scale-free form $\rho_\mathrm{gas}/500 \rho_\mathrm{crit,0}$ and traced over the scaled halocentric radius, $r/r_\mathrm{500}$, indicated by the light blue shaded region, as predicted for the ideal baryonic cluster halos. This shaded region evaluates the five parameters of our model, shown in Table \ref{Parameter table}, at values chosen to correspond to the overdensity $\Delta=500$. This prediction is compared to recent observational gas density fits: \cite{Pratt2023} (the blue dotted line) and \cite{Lyskova2023} (the orange dash-dotted line), with the hydrogen gas density in the latter converted to a gas density.}
    \includegraphics[width=\textwidth]{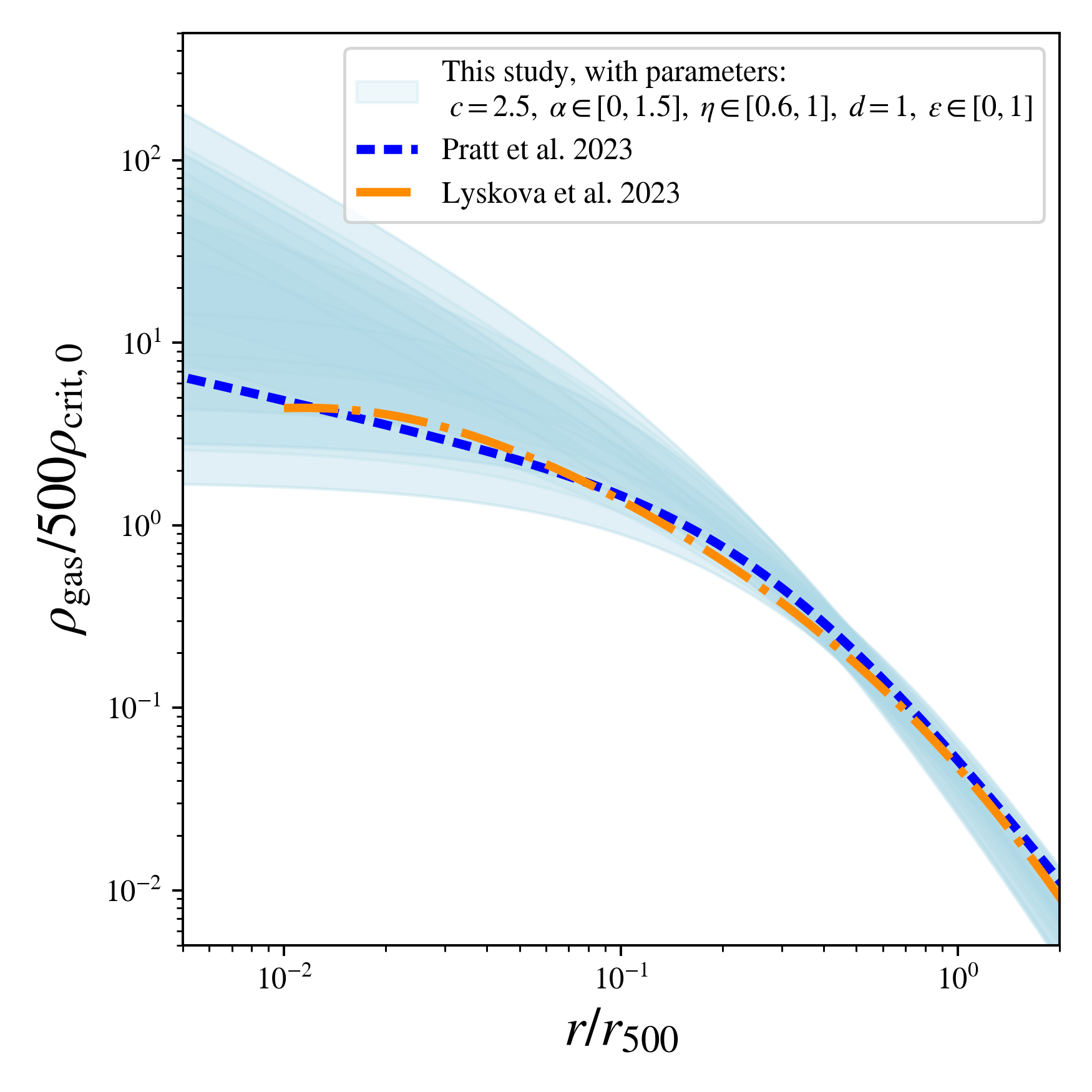}
    \label{Fig - ideal physical baryonic halo profiles}
\end{figure}

\begin{table*}[h!]
\begin{tabular}{ |p{2cm}||p{2.2cm}|p{3cm}|p{2.8cm}|p{2.2cm}|p{3cm}| }
\hline
&  \multicolumn{1}{|c|}{$\boldsymbol{c}$} & \multicolumn{1}{|c|}{$\boldsymbol{\alpha}$} & \multicolumn{1}{|c|}{$\boldsymbol{\eta}$} & \multicolumn{1}{|c|}{$\boldsymbol{d}$} & \multicolumn{1}{|c|}{$\boldsymbol{\varepsilon}$} \\ [1ex]
\hline\hline
\rowcolor{white} \cellcolor{lightgray!30} Definition: &  \textit{Concentration} & \textit{Inner density slope of the dark matter profile} & \textit{Fraction of cosmological baryon content}  & \textit{Dilution} & \textit{Inner density slope of the intracluster gas profile} \\ [3ex] 
\hline
\rowcolor{white} \cellcolor{lightgray!30} Physical values: &  $c=5 \quad (\Delta=200),$ \newline $ c=2.5 \quad (\Delta=500)$ & $\alpha \in [0, 1.5]$ & $\eta \in [0.6, 1]$ & $d=2 \quad (\Delta=200),$ \newline $ d=1 \quad (\Delta=500)$ & $\varepsilon \in [0, 1]$ \\ [3ex] 
\hline
\end{tabular}
\caption{\textcolor{black}{Summary of the five parameters in the ideal baryonic cluster halo model: their symbol, definition and physical values.}}
\label{Parameter table}
\end{table*}

\subsubsection{The ideal baryonic cluster halo profiles}

With this parameterisation of the intracluster gas profile, together with a parameterisation for the underlying halo, we can mathematically model the structural composition of idealised cluster halos. We refer to the structures modelled by these idealised profiles as `ideal baryonic cluster halos'.

Importantly, when modelling the dark matter halo's density profile by Equation \eqref{ideal physical halo profile} (i.e. the ideal physical halo considered in Paper I), we must apply a scalar factor of $(1 - \eta f_\mathrm{b, cos})$ to take into account the gas mass contribution. This scalar factor ensures that the total density of the cluster halo, now considering the gas contribution, still integrates to the normalised virial mass. This results in a slightly adjusted dark matter density profile to model the ideal baryonic cluster halos, of the form:
\begin{equation}\label{ideal physical baryonic halo profile - dark matter}
    \frac{\rho_\mathrm{dm} (s, c, \alpha, \eta)}{\Delta \rho_\mathrm{crit, 0}} = \frac{(1  - \eta f_\mathrm{b, cos}) u(c, \alpha)}{3s^\alpha (1 + cs)^{3-\alpha}},
\end{equation}
where the halo's concentration function, $u(c, \alpha)$, remains defined by Equation \eqref{ideal physical halo concentration function}. In terms of the gas structural parameters detailed, the gas profile to model the ideal baryonic cluster halos takes the form:
\begin{equation}\label{ideal physical baryonic halo profile - baryonic gas}
    \frac{\rho_\mathrm{gas}(s, c, \alpha, \eta, d, \varepsilon)}{\Delta \rho_\mathrm{crit,0}} = \frac{\eta f_\mathrm{b, cos}\mathcal{U}(c, \alpha, d, \varepsilon)}{3s^\varepsilon [1 + \mathcal{C}(c, \alpha, d, \varepsilon) s ]^{3-\varepsilon}},
\end{equation}
where the gas' concentration parameter, $\mathcal{C}(c, \alpha, d, \varepsilon)$, and the gas' generalised concentration function, $\mathcal{U}(c, \alpha, d, \varepsilon)$, are defined in Equations \eqref{baryon concentration parameter} and \eqref{baryon concentration function}, respectively. 

These idealised density profiles are constrained by five parameters --- the underlying dark matter halo's structural parameters: its concentration, $c$, and inner slope, $\alpha$; the gas distribution's structural parameters: its dilution, $d$, and inner slope, $\varepsilon$; and the relative contribution of each component, set by $\eta$, the fraction of cosmological baryon content. As detailed, these five parameters take on physically-motivated values: three are bounded continuously: $\alpha \in [0, 1.5]$, $\eta \in [0.6, 1]$ and $\varepsilon \in [0, 1]$, and the remaining two are set at fixed values, depending on the overdensity: $c=5$ and $d=2$ when $\Delta=200$, and $c=2.5$ and $d=1$ when $\Delta=500$. \textcolor{black}{These parameters and their chosen values are summarised in Table \ref{Parameter table}.}

The resulting parameter space of intracluster gas density profiles predicted in this model is shown in the grey shaded region of Figure \ref{Fig - ideal physical baryonic halo profiles}, with the overdensity is taken as $\Delta=500$. Here, two recent observational constraints on the average intracluster gas density profile are shown in comparison to our model: the best-fit from \cite{Pratt2023} shown by the blue-dotted line, from 93 clusters observed with XMM-\textit{Newton} and fit to a generalised NFW density model; and the best-fit from \cite{Lyskova2023} shown by the orange-dotted line, from 38 clusters observed with eROSITA and fit to a ten-parameter function. Both of these independent observational best-fits are well contained within the predicted parameter space of our model, motivating the use of this analytic model to allow predictions for the observable X-ray and SZ emission properties of galaxy clusters.

\subsection{2.3. \quad Thermodynamic profiles of cluster halos}\label{Section 2.3}

For idealised, spherically symmetric gravitational systems consisting of gas and dark matter components, simple analytical expressions can be used to predict the corresponding thermodynamic emission profiles.

\subsubsection{Hydrostatic equilibrium}

To relate the gravitational mass of a cluster halo to its thermodynamic state, it is typically assumed that the intracluster gas is in a state of hydrostatic equilibrium: such that the thermal pressure exerted by the hot gas is balanced by the halo's gravitational potential. For a cluster halo consisting of dark matter, $\rho_\mathrm{dm}(r)$, and hot gas, $\rho_\mathrm{gas}(r)$, density contributions, the hydrostatic equilibrium state is solved by the differential equation:
\begin{equation}\label{hydrostatic equilibrium condition}
    \frac{\mathrm{d}}{\mathrm{d}r}\left[ \rho_\mathrm{gas}(r) T(r)\right] = - \frac{G\mu m_\mathrm{p}}{k_\mathrm{B}} \frac{\rho_\mathrm{gas}(r) M(r)}{r^2} ,
\end{equation}
where $M(r)$ is the total mass of the halo cluster within halocentric radius $r$, given by integrating the density components of the system:
\begin{equation}\label{total mass (dark matter + baryonic gas)}
     M(r) = 4\pi \int_0 ^r \left[ \rho_\mathrm{dm}(r^\prime) + \rho_\mathrm{gas}(r^\prime) \right] r^\prime{} ^2 \mathrm{d}r^\prime,
\end{equation}
and where $T(r)$ is the equilibrium temperature of the gas, to be solved for. In Equation \eqref{hydrostatic equilibrium condition}, the physical constants involved are: Newton's gravitational constant, $G$, the mean molecular gas weight, $\mu$, the proton mass, $m_\mathrm{p}$ and the Boltzmann constant, $k_\mathrm{B}$. Given profiles for the gas and the dark matter --- i.e. the ideal baryonic cluster halos posited in this investigation --- and some initial or boundary condition on the system, this differential equation can be solved for the gas' equilibrium temperature, $T(r)$, over the radial extent of cluster. 

Analytic studies have circumvented prescribing the form of the gas density profile or any boundary conditions in solving the cluster's temperature distribution, by imposing a polytropic equation of state on the gas \autocite[see, e.g.][]{Komatsu&Seljak2001}. However, observational fits for the gas and temperature profiles of clusters have shown that this polytropic relation fails at large radii \autocite[e.g.][]{DeGrandi&Molendi2002}, whilst these polytropic models fail to predict the rise in gas temperature with halo radii in the central region as consistently found observationally \autocite[e.g.][]{Vikhlinin2006, Sun2009, Lyskova2023}. As such, in this study we will not assume a polytropic relation for the gas, and instead solve for the hydrostatic equilibrium state of the gas by imposing a sensible boundary condition on the gas temperature: $\lim _{r\to\infty} T(r) = 0$; or, equivalently, that the temperature goes to zero at very large cluster radii. This allows the general solution for the gas' equilibrium temperature, solving the hydrostatic state in Equation \eqref{hydrostatic equilibrium condition}, to be expressed as:
\begin{equation}\label{temperature solution}
    T(r) = \frac{G\mu m_\mathrm{p}}{k_\mathrm{B}} \frac{1}{\rho_\mathrm{gas}(r)} \int _r^\infty \frac{M(r^\prime) \rho_\mathrm{gas}(r^\prime) \mathrm{d}r^\prime}{r^\prime{} ^2}.
\end{equation}

To allow this equilibrium temperature to be composed in a scale-free formulation, we can introduce the virial temperature, $T_\mathrm{vir}$, defined as the temperature of a gas distribution, at halocentric radius, $r_\mathrm{vir}$, for a gravitational system in virial and hydrostatic equilibrium, as:
\begin{equation}\label{virial temperature definition}
    T_\mathrm{vir} \equiv \frac{1}{3} \frac{\mu m_\mathrm{p}}{k_\mathrm{B}} \frac{GM_\mathrm{vir}}{r_\mathrm{vir}}.
\end{equation}
Of note, in this definition of $T_\mathrm{vir}$ in Equation \eqref{virial temperature definition}, the prefactor of $1/3$ is not unique: some definitions instead will use a prefactor of $1/2$ instead, with these differences arising due to assumptions in the form of the gas profile deriving the virial parameter. As we will only use $T_\mathrm{vir}$ as a normalisation factor, and remain consistent throughout this analysis, the choice in prefactor is entirely ambiguous and will not impact our predictions. 

This general solution for the equilibrium temperature in Equation \eqref{temperature solution} can then be expressed as a scale-free profile, in a ratio to the virial temperature and as a function of the dimensionless radius, $s$, taking the form:
\begin{equation}\label{scale-free temperature}
    \frac{T(s)}{T_\mathrm{vir}} = 3 \frac{\Delta \rho_\mathrm{crit,0}}{\rho_\mathrm{gas}(s)} \int _s ^\infty \frac{M(s^\prime)}{M_\mathrm{vir}} \frac{\rho_\mathrm{gas}(s^\prime)}{\Delta \rho_\mathrm{crit,0}} \frac{\mathrm{d}s^\prime}{s^\prime{} ^2},
\end{equation}
in terms of the scale-free profiles for the intracluster gas, density $\rho_\mathrm{gas}(s)$, and the total mass, $M(s)$, itself encoding both density constituents comprising the system:
\begin{equation}\label{scale-free total mass (dark matter + baryonic gas)}
    \frac{M(s)}{M_\mathrm{vir}} = 3\int _0 ^s \left[ \frac{\rho_\mathrm{dm}(s^\prime)}{\Delta \rho_\mathrm{crit,0}} + \frac{\rho_\mathrm{gas}(s^\prime)}{\Delta \rho_\mathrm{crit,0}} \right] s^\prime{} ^2 \mathrm{d}s^\prime.
\end{equation}

\textcolor{black}{The corresponding equilibrium pressure, $p$, of the intracluster gas is then related to the equilibrium temperature and density of the gas by the ideal gas law:
\begin{equation}\label{ideal gas law}
    p = \frac{k_\mathrm{B} T}{\mu m_\mathrm{p}} \rho_\mathrm{gas},
\end{equation}
hence the scale-free pressure profile of the gas, $p(s)$, will be given by the profile:
\begin{equation}\label{scale-free pressure}
    \frac{p(s)}{p_\mathrm{vir}} = 3 \int _s ^\infty \frac{M(s^\prime)}{M_\mathrm{vir}} \frac{\rho_\mathrm{gas}(s^\prime)}{f_\mathrm{b, cos} \Delta \rho_\mathrm{crit,0}} \frac{\mathrm{d}s^\prime}{s^\prime{} ^2},
\end{equation}
with the total mass, $M(s)$, still defined by Equation \eqref{scale-free total mass (dark matter + baryonic gas)}, and where the parameter $p_\mathrm{vir}$ is defined as the virial pressure:
\begin{equation}\label{virial pressure definition}
    p_\mathrm{vir} \equiv \frac{k_\mathrm{B}T_\mathrm{vir}}{\mu m_\mathrm{p}} f_\mathrm{b, cos} \Delta \rho_\mathrm{crit,0},
\end{equation}
as the pressure of a halo with temperature $T_\mathrm{vir}$ at halocentric radius $r_\mathrm{vir}$, given the halo is in virial and hydrostatic equilibrium, and with a gas fraction of exactly cosmological baryon content, $f_\mathrm{b, cos}$.}

\subsubsection{The emission-weighted temperature}

In X-ray observations, the temperature of a cluster is typically weighted by the properties of the emitting gas for an average measure of the cluster's temperature. Most commonly, this is the emission-weighted temperature, or simply the X-ray temperature, $T_\mathrm{X}(<r_\mathrm{det})$, as measured when the cluster's temperature is weighted by the gas' emission out to a halocentric radius, called the detection radius, $r_\mathrm{det}$. For a spherically symmetric galaxy cluster, the emission-weighted temperature can be calculated by the integral:
\begin{equation}\label{emission-weighted temperature}
    T_\mathrm{X}(<r_\mathrm{det}) = \frac{\int _0 ^{r_\mathrm{det}}\rho_\mathrm{gas}^2(r) \Lambda (T) T(r)  r^2 \mathrm{d}r}{\int _0 ^{r_\mathrm{det}}  \rho_\mathrm{gas}^2(r) \Lambda (T) r^2 \mathrm{d}r},
\end{equation}
where the emission is taken to be proportional to $\rho_\mathrm{gas}^2(r)\Lambda(T)$, in terms of the cooling function, $\Lambda (T)$, which depends on the emission mechanism that dominates at the cluster's physical temperature. For X-ray emitting hot gas, Bremsstrahlung or `free-free' emission dominates, with X-ray emission proportional to $\rho_\mathrm{gas}^2(r) T^{1/2}(r)$. In this regime, the emission-weighted temperature is given by:
\begin{equation}\label{X-ray emission-weighted temperature}
    T_\mathrm{X}(<r_\mathrm{det}) = \frac{\int _0 ^{r_\mathrm{det}}  \rho_\mathrm{gas}^2(r) T^{3/2}(r) r^2 \mathrm{d}r}{\int _0 ^{r_\mathrm{det}} \rho_\mathrm{gas}^2(r) T^{1/2}(r) r^2 \mathrm{d}r},
\end{equation}
which is accessible in X-ray cluster observations, given fits for a cluster's radial gas density and temperature profiles out to some detection radius, usually limited to $r_{500}$. Formulating this expression in terms of the dimensionless radius, $s$, and a corresponding dimensionless detection radius, $s_\mathrm{det}$, the scale-free formulation of this X-ray temperature is:
\begin{equation}\label{scale-free X-ray emission-weighted temperature}
    \frac{T_\mathrm{X}(<s_\mathrm{det})}{T_\mathrm{vir}} = \frac{\int _0 ^{s_\mathrm{det}}  \left[\frac{\rho_\mathrm{gas}(s)}{\Delta \rho_\mathrm{crit,0}}\right]^2 \left[\frac{T(s)}{T_\mathrm{vir}}\right]^{3/2} s^2 \mathrm{d}s}{\int _0 ^{s_\mathrm{det}}  \left[\frac{\rho_\mathrm{gas}(s)}{\Delta \rho_\mathrm{crit,0}}\right]^2 \left[\frac{T(s)}{T_\mathrm{vir}}\right]^{1/2} s^2 \mathrm{d}s},
\end{equation}
as weighted by the scale-free profiles for the cluster's gas density and temperature. 

\subsubsection{The mean gas mass-weighted temperature}

Often, rather than measuring the emission-weighted temperature, observational surveys can instead measure a mean gas mass-weighted temperature, $T_\mathrm{m_g}(<r_\mathrm{det})$, similarly measured out to a halocentric radius of $r_\mathrm{det}$. For a spherically symmetric mass distribution, this observable is recovered by weighting the temperature by the gas density profile, such that:
\begin{equation}\label{mean gas mass-weighted temperature}
    T_\mathrm{m_g} (<r_\mathrm{det}) = \frac{\int _0 ^{r_\mathrm{det}}  \rho_\mathrm{gas}(r) T(r) r^2 \mathrm{d}r}{\int _0 ^{r_\mathrm{det}}  \rho_\mathrm{gas}(r) r^2 \mathrm{d}r}.
\end{equation}
This weighted temperature is similarly accessible in X-ray observations, given fits to the cluster's gas density and temperature. Due to its weighting by the mean gas mass, which should be proportional to the total halo mass, this observable is often preferred in X-ray scaling relations, as a tight proportionality to the halo mass is expected. In terms of our scale-free framework, this temperature measure is:
\begin{equation}\label{scale-free mean gas mass-weighted temperature}
    \frac{T_\mathrm{m_g} (<s_\mathrm{det})}{T_\mathrm{vir}} = \frac{\int _0 ^{s_\mathrm{det}}  \frac{\rho_\mathrm{gas}(s)}{\Delta \rho_\mathrm{crit,0}} \frac{T(s)}{T_\mathrm{vir}} s^2 \mathrm{d}s}{\int _0 ^{s_\mathrm{det}}  \frac{\rho_\mathrm{gas}(s)}{\Delta \rho_\mathrm{crit,0}} s^2 \mathrm{d}s},
\end{equation}
as weighted by the scale-free profiles within the dimensionless detection radius, $s_\mathrm{det}$. 

\subsubsection{The Sunyaev-Zeldovich signal}

The SZ effect is known to induce a frequency-dependent temperature shift, $\Delta T_\mathrm{SZE}$, on the CMB temperature, $T_\mathrm{CMB}$, of the form \autocite{SZ1970, SZ1972}:
\begin{equation}\label{SZ effect}
    \frac{\Delta T_\mathrm{SZE}}{T_\mathrm{CMB}} = f(\nu ) \cdot y,
\end{equation}
where $f(\nu)$ encodes the frequency dependence and $y$ is the Compton parameter. The SZ signal for a galaxy cluster is encoded in this Compton parameter, which measures the cluster's electron pressure, $p_\mathrm{e}$, integrated inside some volume, $\mathcal{V}$, as:
\begin{equation}\label{Compton parameter definition}
    y \equiv \frac{\sigma_\mathrm{T}}{m_\mathrm{e}c_\gamma^2} \int _{\mathcal{V}} p_\mathrm{e} \mathrm{d}\mathcal{V},
\end{equation}
usually integrated along the line of sight. In Equation \eqref{Compton parameter definition}, the physical constants are: the Thompson cross-section, $\sigma_\mathrm{T}$, the speed of light, $c_\gamma$ \footnote{The speed of light is denoted with a subscript, as $c_\gamma$, to avoid confusion with the halo's concentration, denoted $c$.}, and the electron mass, $m_\mathrm{e}$. The electron pressure is related to the intracluster gas pressure, $p$, by the ratio of the mean molecular weight, $\mu$, to the mean molecular weight of electrons, $\mu_\mathrm{e}$, and with this gas pressure related to the gas' density and temperature by the ideal gas law, Equation \eqref{ideal gas law}, such that the electron pressure can be expressed as:
\begin{equation}\label{electron pressure}
    p_\mathrm{e} = \frac{\mu}{\mu_\mathrm{e}} p = \frac{k_\mathrm{B} T}{\mu_\mathrm{e} m_\mathrm{p}} \rho_\mathrm{gas}.
\end{equation}
As such, the Compton parameter can be equivalently defined by the volume integral:
\begin{equation}\label{Compton parameter, gas variables}
    y \equiv \frac{k_\mathrm{B} \sigma_\mathrm{T}}{m_\mathrm{e} c_\gamma ^2 \mu _\mathrm{e} m_\mathrm{p}} \int _{\mathcal{V}} \rho_\mathrm{gas} T \mathrm{d} \mathcal{V}. 
\end{equation}

In observational surveys, the SZ signal for a galaxy cluster can be measured by calculating the Compton parameter as in Equation \eqref{Compton parameter, gas variables} when integrated over some well-defined volume: typically, this is either a spherical volume, by integrating fits for the cluster's 3-dimensional gas and temperature profiles, or in a cylindrical volume, by integrating these fits along the line of sight direction within some 2-dimensional projected radius, $R_\mathrm{ap}$, known as the aperture radius. In these integral expressions, we will consistently notate 3-dimensional halo radii with a lower-case $r$, and 2-dimensional projected radii with an upper-case $R$, inclusive of all subscripts. 

For a spherically-integrated, $Y_\mathrm{sph} (<r_\mathrm{det})$, Compton parameter, measured within the sphere of halocentric radius of $r_\mathrm{det}$, the associated SZ signal is given by:
\begin{equation}\label{spherical SZ signal}
    Y_\mathrm{sph} (<r_\mathrm{det}) = \frac{4\pi k_\mathrm{B} \sigma_\mathrm{T} }{m_\mathrm{e} c_\gamma^2 \mu_\mathrm{e} m_\mathrm{p}} \int _0 ^{r_\mathrm{det}} \rho_\mathrm{gas}(r) T(r) r^2 \mathrm{d} r.
\end{equation}
Similarly, for a cylindrically-integrated, $Y_\mathrm{cyl}(<R_\mathrm{ap})$, Compton parameter, measured along the line of sight direction within an aperture radius, $R_\mathrm{ap}$, the associated SZ signal will be:
\begin{equation}\label{cylindrical SZ signal}
    Y_\mathrm{cyl}(<R_\mathrm{ap}, r_\mathrm{b}) = \frac{4\pi k_\mathrm{B} \sigma_\mathrm{T}}{m_\mathrm{e} c_\gamma ^2 \mu_\mathrm{e} m_\mathrm{p}} \int_0 ^{R_\mathrm{ap}} R \mathrm{d} R \Biggl[\int _R ^{r_\mathrm{b}} \frac{\rho_\mathrm{gas}(r) T(r) r\mathrm{d}r}{\sqrt{r^2 - R^2}}\Biggr];
\end{equation}
or, equivalently, by subtracting from the total spherically-integrated parameter at the halocentric cluster boundary, $r_\mathrm{b}$, as \autocite{Arnaud2010}:
\begin{equation}\label{cylindrical SZ signal - simplified}
\begin{aligned}
    Y_\mathrm{cyl}(<R_\mathrm{ap}, r_\mathrm{b}) &= Y_\mathrm{sph}(<r_\mathrm{b}) \\
    & - \frac{4\pi k_\mathrm{B} \sigma_\mathrm{T}}{m_\mathrm{e} c_\gamma ^2 \mu_\mathrm{e} m_\mathrm{p}} \int_{R_\mathrm{ap}}^{r_\mathrm{b}} \rho_\mathrm{gas}(r) T(r) r\sqrt{r^2 - R_\mathrm{ap}^2} \mathrm{d}r.
\end{aligned}
\end{equation}
In Equations \eqref{cylindrical SZ signal} and \eqref{cylindrical SZ signal - simplified}, this halocentric cluster boundary, $r_\mathrm{b}$, is taken to occur at the cluster's accretion shock, beyond which the electron pressure is expected to fall to the ambient pressure of the intergalactic medium, thus truncating the cluster's SZ signal. This cluster boundary is usually taken as $r_\mathrm{b} \simeq 5 r_{500}$, as predicted in hydrodynamic simulations \autocite[see, e.g.][]{Arnaud2010}. 

When the SZ effect is integrated over some volume, it becomes an important observational probe, as the integrated Compton parameter will be proportional to the number of electrons inside the region, and so proportional to the cluster's mass. To capture these SZ observables in a scale-free formalism, we introduce the virial Compton parameter, $Y_\mathrm{vir}$, defined as the spherically-integrated value enclosed by the halocentric radius $r_\mathrm{vir}$, as:
\begin{equation}\label{virial SZ definition}
    Y_\mathrm{vir} \equiv \frac{4}{3} \pi r_\mathrm{vir}^3  \frac{ \sigma_\mathrm{T}}{m_\mathrm{e} c_\gamma ^2 } \frac{\mu }{\mu_\mathrm{e}} p_\mathrm{vir},
\end{equation}
\textcolor{black}{where $p_\mathrm{vir}$ is the virial pressure, defined in Equation \eqref{virial pressure definition}.} Taking this definition of $Y_\mathrm{vir}$, the scale-free spherically-integrated Compton parameter, measured within a dimensionless detection radius, $s_\mathrm{det}$, is given by the profile:
\begin{equation}\label{scale-free spherical SZ signal}
    \frac{Y_\mathrm{sph}(<s_\mathrm{det})}{Y_\mathrm{vir}} = 3\int _0 ^{s_\mathrm{det}} \frac{\rho_\mathrm{gas}(s)}{f_\mathrm{b, cos} \Delta \rho_\mathrm{crit,0}} \frac{T(s)}{T_\mathrm{vir}} s^2 \mathrm{d}s.
\end{equation}
To devise an expression for the scale-free cylindrically-integrated Compton parameter, we must define a dimensionless projected radius scale, denoted by $S$, as the ratio of the projected radius to the virial radius, as:
\begin{equation}\label{dimensionless projected radius definition}
    S \equiv \frac{R}{r_\mathrm{vir}}.
\end{equation}
As such, in terms of a dimensionless aperture radius, $S_\mathrm{ap}$, the scale-free cylindrically-integrated Compton parameter is given by the profile:
\begin{equation}\label{scale-free cylindrical SZ signal}
\begin{aligned}
    \frac{Y_\mathrm{cyl}(<S_\mathrm{ap}, s_\mathrm{b})}{Y_\mathrm{vir}} &= \frac{Y_\mathrm{sph}(<s_\mathrm{b})}{Y_\mathrm{vir}} \\
    & - 3\int _{S_\mathrm{ap}}^{s_\mathrm{b}} \frac{\rho_\mathrm{gas} (s)}{f_\mathrm{b, cos} \Delta \rho_\mathrm{crit,0}} \frac{T(s)}{T_\mathrm{vir}} s \sqrt{s^2 - S_\mathrm{ap}^2} \mathrm{d} s,
\end{aligned}
\end{equation}

as subtracted from the total, scale-free spherically-integrated signal from Equation \eqref{scale-free spherical SZ signal}, and where all three-dimensional, dimensionless halo radii are denoted by a lower-case $s$, and all two-dimensional, dimensionless projected radii are denoted by an upper-case $S$, inclusive of all subscripts. 

In this scale-free composition, $s_\mathrm{b} \equiv r_\mathrm{b}/r_\mathrm{vir}$ is now the overdensity-dependent cluster boundary, requiring its specification in units of $r_{200}$ or $r_{500}$, when $\Delta$ is fixed. Taking the predicted value $r_\mathrm{b} = 5 r_{500}$ when $\Delta=500$, we will assume this is equivalently approximated to $r_\mathrm{b} = 2.5 r_{200}$ when $\Delta=200$, approximately consistent with numerical predictions \autocite[e.g.][]{Lau2015}.

\subsection{2.4. \quad Predictions for the X-ray and SZ scaling relations}

With this scale-free framework for X-ray and SZ observables, we now seek to construct an analytical correspondence between these observables and the halo mass, in the form of scaling relations. 

\subsubsection{Dimensional analysis}

For this derivation, we use the method of dimensional analysis. This technique allows us to link the halo's mass, $M_\mathrm{halo}$ (measured in $\mathrm{M}_\odot$), to the temperature observables, $T_\mathrm{X}$, $T_\mathrm{m_g}$ (measured in $\mathrm{K}$), and the SZ observables, $Y_\mathrm{sph}$, $Y_\mathrm{cyl}$ (measured in $\mathrm{Mpc}^2$), by some combination of physical constants that ensure dimensionality in each scaling relation, and by the introduction of some dimensionless prefactor, denoted $\mathcal{A}$. 

For the temperature observables, we can reasonably assume that the correspondence contains the physical constants: $G$ (measured in $\mathrm{Mpc \, km^2\,s^{-2} \, \mathrm{M}_\odot ^{-1}}$), $\rho_\mathrm{crit,0}$ (measured in $\mathrm{M_\odot \, Mpc^{-3}}$), $k_\mathrm{B}$ (measured in $\mathrm{M_\odot \, km^2 \, s^{-2} \, K^{-1}}$), and $m_\mathrm{p}$ (measured in $\mathrm{M}_\odot$), such that, by dimensionality, this relation will take the form: 
\begin{equation}\label{dimensional analysis}
    M_\mathrm{halo} = \mathcal{A} \cdot \sqrt{\frac{1}{\rho_\mathrm{crit,0}} \left[\frac{k_\mathrm{B}}{ m_\mathrm{p} G}\right] ^3} \cdot T_\mathrm{X}^{3/2},
\end{equation}
where the temperature observable is either $T_\mathrm{X}$, as shown, or $T_\mathrm{m_g}$; this correspondence is interchangeable. For constraints to be placed on this relationship, the dimensionless prefactor, $\mathcal{A}$, must be predicted and constrained. For this, we can use the virial theorem, to relate the halo's virial mass to its virial temperature, by utilising the definitions in Equations \eqref{virial mass definition} and \eqref{virial temperature definition}, whereby:
\begin{equation}\label{virial mass - temperature relation}
    M_\mathrm{vir} = \frac{9}{2} \sqrt{\frac{1}{\pi \Delta \rho_\mathrm{crit,0}} \left[\frac{k_\mathrm{B}}{G \mu m_\mathrm{p} }\right] ^3} \cdot T_\mathrm{vir} ^{3/2},
\end{equation}
with this above relationship encoding the same dimensional form as Equation \eqref{dimensional analysis}. By taking the virial mass, $M_\mathrm{vir}$, as an approximation to the halo's mass, in either convention $M_{200}$ or $M_{500}$, we can use the form of this correspondence to recover the desired scaling relation, such that:
\begin{equation}\label{virial mass bound TX}
    M_\mathrm{vir} = \frac{9}{2} \sqrt{\frac{1}{\pi \Delta \rho_\mathrm{crit,0}} \left[\frac{k_\mathrm{B}}{G \mu m_\mathrm{p}}\right] ^3} \cdot \left[\frac{T_\mathrm{X}}{\tau_1}\right]^{3/2},
\end{equation}
in terms of the emission-weighted temperature, $T_\mathrm{X}$, by the introduction of a dimensionless parameter, denoted $\tau_1$. Equivalently, this correspondence can be constructed in terms of the mean gas mass-weighted temperature, $T_\mathrm{m_g}$, by the introduction of a similar dimensionless parameter, $\tau_2$, such that:
\begin{equation}\label{virial mass bound Tmg}
    M_\mathrm{vir} = \frac{9}{2} \sqrt{\frac{1}{\pi \Delta \rho_\mathrm{crit,0}} \left[\frac{k_\mathrm{B}}{G \mu m_\mathrm{p}}\right] ^3} \cdot \left[\frac{T_\mathrm{m_g}}{\tau_2}\right]^{3/2}.
\end{equation}
In these scaling relations above, the dimensionless parameters $\tau_1$ and $\tau_2$ are each defined as:
\begin{equation}\label{tau definition}
    \tau_1 \equiv \frac{T_\mathrm{X}(<r_\mathrm{det})}{T_\mathrm{vir}}, \quad \tau_2 \equiv \frac{T_\mathrm{m_g}(<r_\mathrm{det})}{T_\mathrm{vir}},
\end{equation}
as the scale-free form of the weighted temperatures, $T_\mathrm{X}$ and $T_\mathrm{m_g}$, respectively, when measured within some detection radius, $r_\mathrm{det}$. Each of these parameters can then be analytically predicted, by Equations \eqref{scale-free X-ray emission-weighted temperature} and \eqref{scale-free mean gas mass-weighted temperature}.

Using this same technique, we can predict the scaling relations between the halo mass and the SZ observables. When considering the correspondence between the virial mass and the virial Compton parameter, defined in Equation \eqref{virial SZ definition}, these scaling relations can be recovered by the introduction of similarly defined dimensionless parameters. For the spherically-integrated Compton parameter, the scaling takes the form:
\begin{equation}\label{virial mass bound SZ, spherical}
    M_\mathrm{vir} = \left[\frac{81}{4 \pi \Delta \rho_\mathrm{crit,0}} \left(\frac{\mu_\mathrm{e} m_\mathrm{e} c_\gamma ^2 }{\mu \sigma_\mathrm{T} f_\mathrm{b, cos} G}\right)^3\right]^{1/5} \cdot \left[\frac{Y_\mathrm{sph}}{\zeta_1}\right] ^{3/5},
\end{equation}
and for the cylindrically-integrated Compton parameter, the form:
\begin{equation}\label{virial mass bound SZ, cylindrical}
    M_\mathrm{vir} = \left[\frac{81}{4 \pi \Delta \rho_\mathrm{crit,0}} \left(\frac{\mu_\mathrm{e} m_\mathrm{e} c_\gamma ^2 }{\mu \sigma_\mathrm{T} f_\mathrm{b, cos} G}\right)^3\right]^{1/5} \cdot \left[\frac{Y_\mathrm{cyl}}{\zeta_2}\right] ^{3/5}.
\end{equation}
In these relations, the dimensionless parameters $\zeta_1$ and $\zeta_2$ are each defined as:
\begin{equation}\label{zeta definition}
    \zeta_1 \equiv \frac{Y_\mathrm{sph}(<r_\mathrm{det})}{Y_\mathrm{vir}}, \quad \zeta_2 \equiv \frac{Y_\mathrm{cyl}(<R_\mathrm{ap})}{Y_\mathrm{vir}},
\end{equation}
as the scale-free form of the SZ observables, each which can be analytically predicted in Equations \eqref{scale-free spherical SZ signal} and \eqref{scale-free cylindrical SZ signal}, respectively.

\subsubsection{Constraining the scaling relations}

When modelling each of these dimensionless parameters, $\tau_1$, $\tau_2$, $\zeta_1$ and $\zeta_2$, continuously over the parameter space of ideal baryonic cluster halos, as detailed in Table \ref{Parameter table}, each of these dimensionless parameters will be bounded within some interval, when measured within a detection radius, $r_\mathrm{det}$, or in the case of $Y_\mathrm{cyl}$, within an aperture radius, $R_\mathrm{ap}$. From these bounds, the scaling relations in Equations \eqref{virial mass bound TX}, \eqref{virial mass bound Tmg}, \eqref{virial mass bound SZ, spherical} and \eqref{virial mass bound SZ, cylindrical} will each be constrained, within a corresponding minimum and maximum proportionality. 

Evaluating the physical constants in Equation \eqref{virial mass bound TX} and assuming a mean molecular weight $\mu = 0.60$, the halo's X-ray temperature scaling relation $M_\mathrm{vir} - T_\mathrm{X}$ reduces to the form:
\begin{equation}\label{temperature scaling relation - evaluated}
    \frac{M_\mathrm{vir}}{\mathrm{M}_\odot} = \frac{2.757\times 10^4}{\tau _\mathrm{1}^{3/2} \Delta ^{1/2} h } \left[\frac{T_\mathrm{X}}{\mathrm{K}}\right]^{3/2},
\end{equation}
which will have the same numerical form as the $M_\mathrm{vir} - T_\mathrm{m_g}$ scaling relation, Equation \eqref{virial mass bound Tmg}, when replacing $\tau_1$ with $\tau_2$. Here, $h$ is the Hubble parameter, $h \equiv H_0 / 100 \mathrm{km\, s^{-1} Mpc^{-1}}$, for $H_0$ the Hubble constant. In this analysis, we take the value $h=0.6751$ \autocite{Planck2016}, neglecting its uncertainty, as this will be tiny compared to the anticipated bounds in the scaling relations. In Equation \eqref{temperature scaling relation - evaluated}, the values for the overdensity, $\Delta$, must be chosen to specify the virial mass approximation. Taking the convention $\Delta=200$ in Equation \eqref{temperature scaling relation - evaluated}, the $M_{200} - T_\mathrm{X}$ scaling relation takes the form:
\begin{equation}\label{M200 TX mass bound}
    \frac{M_{200}}{\mathrm{M}_\odot} = \frac{2.888\times 10^3}{\tau _\mathrm{1}^{3/2} } \left[\frac{T_\mathrm{X}}{\mathrm{K}}\right]^{3/2},
\end{equation}
and similarly, in the convention $\Delta=500$, the $M_{500} - T_\mathrm{X}$ scaling relation takes the form:
\begin{equation}\label{M500 TX mass bound}
    \frac{M_{500}}{\mathrm{M}_\odot} = \frac{1.826\times 10^3}{\tau _\mathrm{1}^{3/2} } \left[\frac{T_\mathrm{X}}{\mathrm{K}}\right]^{3/2},
\end{equation}
each dependent on constraints derived for the parameter $\tau_1$. The analogous scaling relations $M_{200} - T_\mathrm{m_g}$ and $M_{500} - T_\mathrm{m_g}$ will each take on the same numeric form, but again will instead depend on the constraints derived for $\tau_2$, rather than $\tau_1$. 

In complete analogy, the SZ scaling relations are determined by evaluating the physical constants in Equation \eqref{virial mass bound SZ, spherical}, assuming a mean molecular weight $\mu = 0.60$, and a mean molecular weight of electrons $\mu_\mathrm{e} = 1.148$, and taking the integrated Compton parameters in units of $\mathrm{Mpc}^2$. As such, the $M_\mathrm{vir} - Y_\mathrm{sph}$ scaling relation takes the form:
\begin{equation}\label{SZ scaling relation - evaluated}
    \frac{M_\mathrm{vir}}{\mathrm{M}_\odot} = \frac{6.388\times 10^{17}}{\zeta_1^{3/5} \Delta ^{1/5} h^{2/5} } \left[\frac{Y_\mathrm{sph}}{\mathrm{Mpc}^2}\right]^{3/5},
\end{equation}
which will have the same numerical form as the $M_\mathrm{vir} - Y_\mathrm{cyl}$ scaling relation, Equation \eqref{virial mass bound SZ, cylindrical}, when replacing $\zeta_1$ with $\zeta_2$. Evaluating this scaling relation for chosen values of $\Delta$, the $M_{200} - Y_\mathrm{sph}$ scaling relation takes the form:
\begin{equation}\label{M200 SZ spherical mass bound}
    \frac{M_{200}}{\mathrm{M}_\odot} = \frac{2.591\times 10^{17}}{\zeta_1^{3/5} } \left[\frac{Y_\mathrm{sph}}{\mathrm{Mpc}^2}\right]^{3/5},
\end{equation}
and the $M_{500} - Y_\mathrm{sph}$ scaling relation, the form:
\begin{equation}\label{M500 SZ spherical mass bound}
    \frac{M_{500}}{\mathrm{M}_\odot} = \frac{2.157\times 10^{17}}{\zeta_1^{3/5} } \left[\frac{Y_\mathrm{sph}}{\mathrm{Mpc}^2}\right]^{3/5},
\end{equation}
as dependent on constraints for $\zeta_1$; and analogously, the scaling relations $M_{200} - Y_\mathrm{cyl}$ and $M_{500} - Y_\mathrm{cyl}$ will take the same form, but instead depend on constraints for $\zeta_2$. 

\subsubsection{Regimes to bound the X-ray and SZ scaling relations}

When deriving these anticipated bounds in the dimensionless parameters, $\tau_1$, $\tau_2$, $\zeta_1$ and $\zeta_2$, the overdensity-dependent parameters --- the concentration, $c$, the dilution, $d$, and the SZ cluster boundary, in the scale-free form $s_\mathrm{b} \equiv r_\mathrm{b}/r_\mathrm{vir}$ --- must each be specified at fixed values. The scale-free detection radius, $s_\mathrm{det} \equiv r_\mathrm{det}/r_\mathrm{vir}$, and aperture radius, $S_\mathrm{ap} \equiv R_\mathrm{ap}/r_\mathrm{vir}$, in real surveys depend on observational contingencies: as they are intrinsically dependent on the available data within a given cluster. Generally, fits for the radial gas profiles are limited to within $r_{500}$, and so $r_\mathrm{det} = r_{500}$ and $R_\mathrm{ap} = r_{500}$ are the most common choice in the literature. As such, we will adopt these values when $\Delta=500$, and when $\Delta=200$ we will adopt the values $r_\mathrm{det} = 0.5r_{200}$ and $R_\mathrm{ap} = 0.5r_{200}$ in rough likeness. 

Given these choices in fixed parameters, the four scaling relations will be evaluated in two distinct regimes, at each overdensity, as summarised below:
\begin{enumerate}
    \item The $M_{200} - T_\mathrm{X}$, $M_{200} - T_\mathrm{m_g}$, $M_{200} - Y_\mathrm{sph}$ scaling relations: \\
    with parameters: $r_\mathrm{det} = 0.5 r_{200}$, $c=5$, $d=2$.

    \item The $M_{200} - Y_\mathrm{cyl}$ scaling relation: \\
    with parameters: $R_\mathrm{ap} = 0.5 r_{200}$, $r_\mathrm{b} = 2.5 r_{200}$, $c=5$, $d=2$.

    \item The $M_{500} - T_\mathrm{X}$, $M_{500} - T_\mathrm{m_g}$, $M_{500} - Y_\mathrm{sph}$ scaling relations: \\
    with parameters: $r_\mathrm{det} = r_{500}$, $c=2.5$, $d=1$.

    \item The $M_{500} - Y_\mathrm{cyl}$ scaling relation: \\
    with parameters: $R_\mathrm{ap} = r_{500}$, $r_\mathrm{b} = 5 r_{500}$, $c=2.5$, $d=1$.
\end{enumerate}
For further clarity, these regimes and their associated parameters are detailed in Tables \ref{Table - tau bounds} and \ref{Table - zeta bounds} in the following section, which show the bounds devised for the dimensionless parameters corresponding to each of these scaling relations, in each of these regimes. 

\section{3. \quad Analysis}
\vspace{2mm}
\subsection{3.1. \quad Analytical profiles for the ideal baryonic cluster halos}

To derive and constrain the dimensionless parameters $\tau_1$, $\tau_2$, $\zeta_1$ and $\zeta_2$, in terms of the outlined parameter space within each of the desired regimes, we will model the dark matter and intracluster gas density profiles with the ideal baryonic cluster halo model, as per Equations \eqref{ideal physical baryonic halo profile - dark matter} and \eqref{ideal physical baryonic halo profile - baryonic gas}, respectively. With this structural model, we can analytically predict and numerically trace the associated scale-free X-ray and SZ emission profiles, and hence the associated dimensionless parameters, by substituting these idealised density forms into the equations for each emission observable, as detailed in Section 2.3. 

\subsubsection{The temperature profile of the ideal baryonic cluster halos}

\begin{figure*}[h!]
    \centering
    \includegraphics[width=\textwidth]{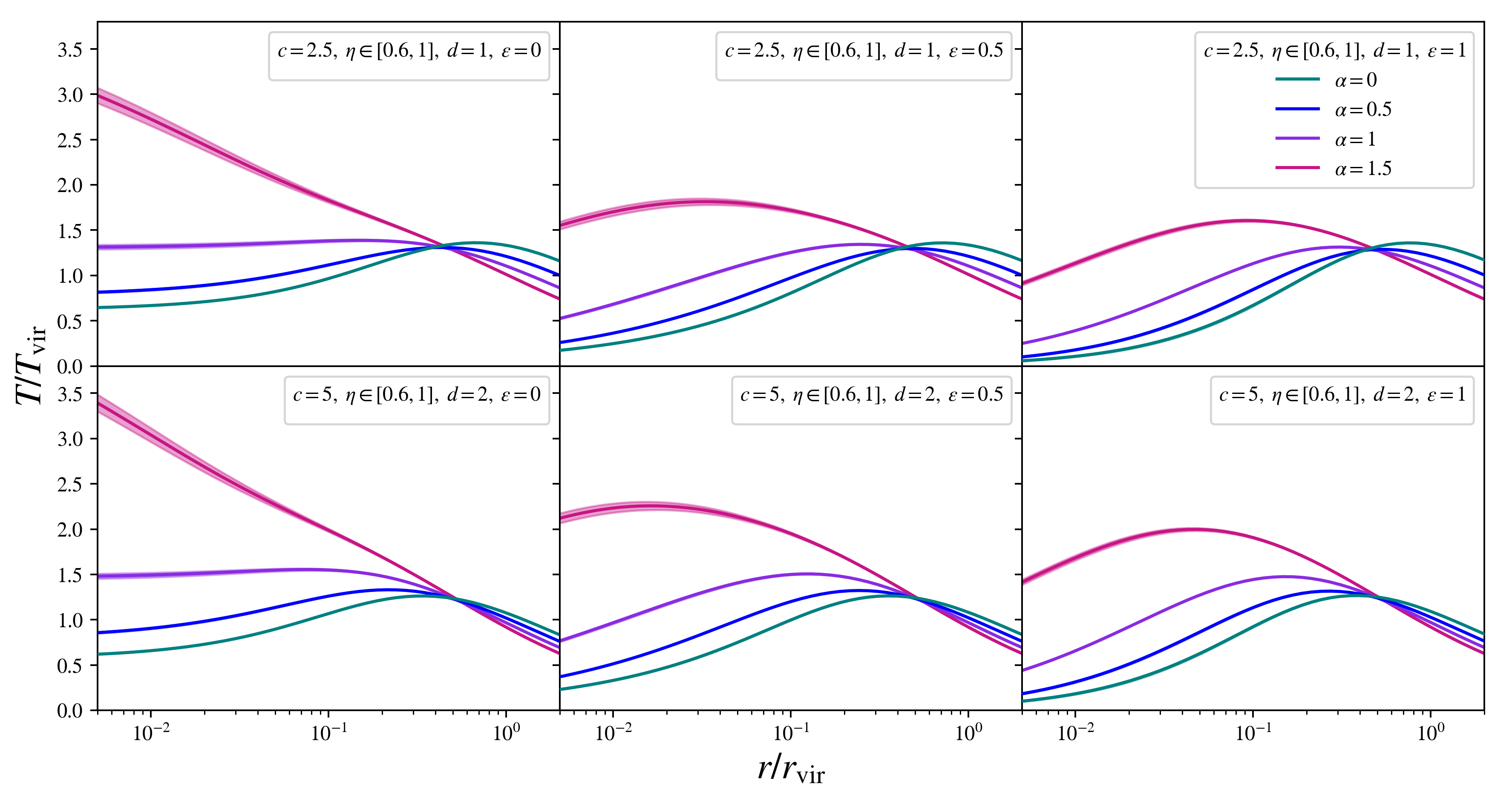}
    \caption{The temperature profiles for the ideal baryonic cluster halos, in scale-free form $T/T_\mathrm{vir}$, traced over the scaled halocentric radius, $r/r_\mathrm{vir}$. Each row varies the halo concentration, $c$, and the dilution, $d$, and each column varies the gas inner slope, $\varepsilon$. Within each box, each colour varies the halo inner slope, $\alpha$, with the solid coloured lines tracing a fraction of cosmological baryon content of $\eta = 0.8$, and the shaded colour region around each solid line (not visible for all curves) tracing this value continuously between $\eta=0.6$ and $\eta = 1$.}
    \label{Fig - temperature profiles}
\end{figure*}

\begin{figure}[h!]
    \centering
    \caption{
    \textcolor{black}{The equilibrium temperature and pressure profiles, in scale-free form $T/T_{500}$ and $p/p_{500}$, shown in the top and bottom panels, respectively, each traced over the scaled halocentric radius, $r/r_\mathrm{500}$, indicated by the light blue shaded regions, as predicted for the ideal baryonic cluster halos. These shaded regions evaluate the five parameters of our model, from Table \ref{Parameter table}, at values chosen to correspond to the overdensity $\Delta=500$. These predictions are compared to recent observational fits for the temperature profile of galaxy clusters, from \cite{Ghirardini2019}, for samples of cool core clusters (the blue dotted line, in the top panel) and non-cool core clusters (the orange dash-dotted line, in the top panel), as well as to the universal gas pressure profile from \cite{Arnaud2010} (the purple dotted line, in the bottom panel).}}
    \includegraphics[width=\textwidth]{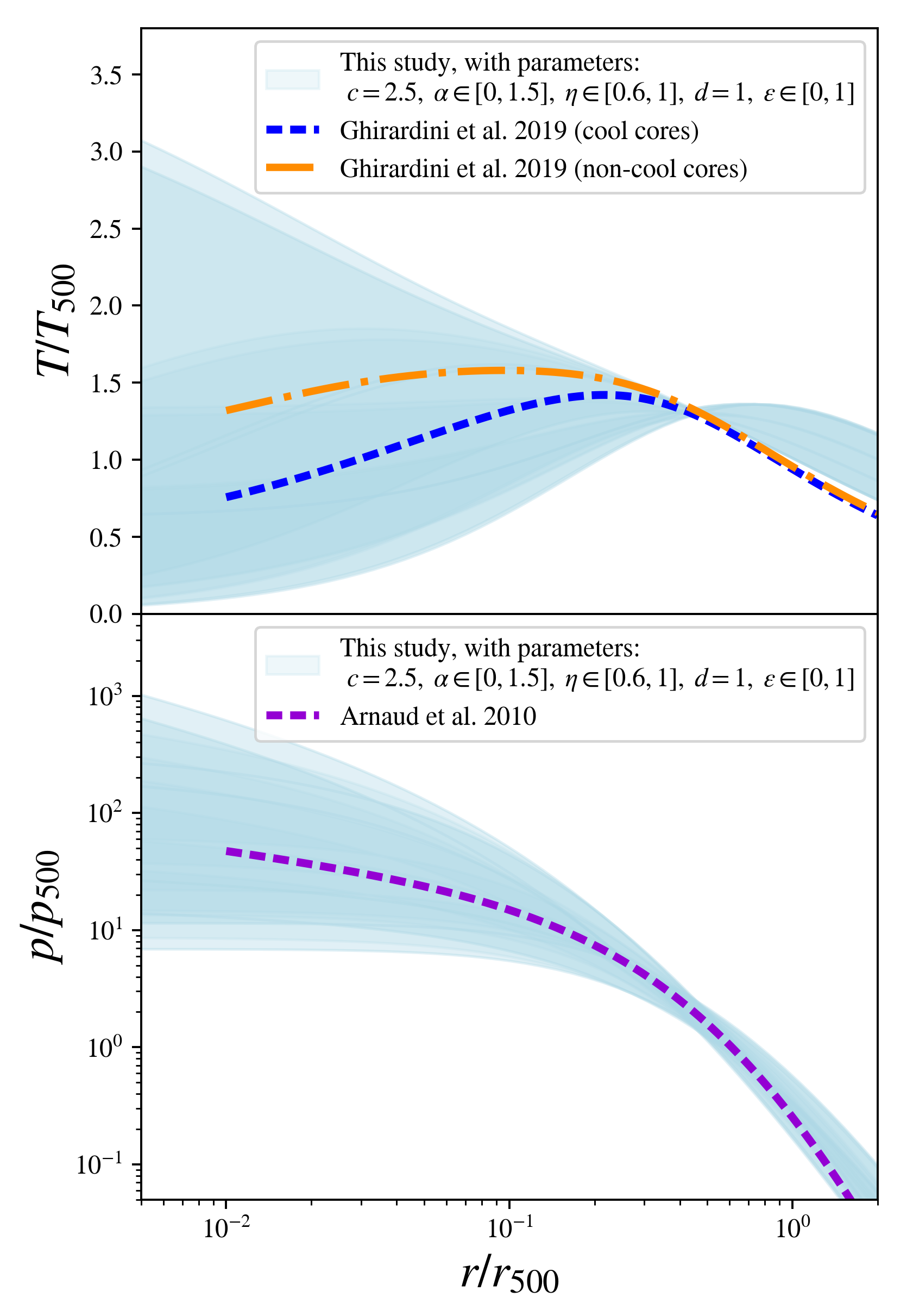}
    \label{Fig - observational comparisons}
\end{figure}

Each of the cluster emission observables depend on the equilibrium temperature profile of the intracluster gas, which is derived from the general solution to the hydrostatic equilibrium state, Equation \eqref{scale-free temperature}. For the ideal baryonic cluster halos, substituting the corresponding density profiles into this solution, the associated scale-free temperature profile takes the analytic form:

\begin{equation}\label{ideal physical baryonic halo temperature}
	\frac{T(s, c, \alpha, \eta, d, \varepsilon)}{T_\mathrm{vir}} = 3s^\varepsilon \left[1 + \mathcal{C}(c, \alpha, d, \varepsilon)s\right]^{3-\varepsilon} \cdot \mathcal{I}(s, c, \alpha, \eta, d, \varepsilon),
\end{equation}
where, for notation simplicity, we define $\mathcal{I}(s, c, \alpha, \eta, d, \varepsilon)$ as the integral function:
\begin{equation}\label{temperature integral function}
\begin{aligned}
    \mathcal{I}(s, c, \alpha, \eta, d, \varepsilon)&\equiv \int _s ^\infty   \frac{ \mathrm{d}s^\prime \Biggl\{  (1 - \eta f_\mathrm{b, cos})u(c, \alpha) \cdot \int _0 ^{s^\prime} \frac{s^\prime{}^\prime{} ^{2-\alpha} \mathrm{d}s^\prime{}^\prime}{(1 + cs^\prime{}^\prime )^{3-\alpha}}   }{s^\prime{}^{2 + \varepsilon} \left[1 + \mathcal{C}(c, \alpha, d, \varepsilon)s^\prime \right]^{3 - \varepsilon}} \\
    & \hspace{-15mm} + \quad \eta f_\mathrm{b, cos} \mathcal{U}(c, \alpha, d, \varepsilon) \cdot \int _0 ^{s^\prime} \frac{s^\prime{}^\prime{}^{2-\varepsilon} \mathrm{d}s^\prime{}^\prime}{\left[1 + \mathcal{C}(c, \alpha, d, \varepsilon)s^\prime{}^\prime \right]^{3-\varepsilon}} \Biggr\}.
\end{aligned}
\end{equation}
These scale-free temperature profiles, $T/T_\mathrm{vir}$, are traced in Figure \ref{Fig - temperature profiles}, as a function of the dimensionless halocentric radius, $s \equiv r/r_\mathrm{vir}$. 
These panels encompass the variation of the halo's temperature within the desired parameter space, for $\alpha \in [0, 1.5]$, $\eta \in [0.6, 1]$ and $\varepsilon \in [0, 1]$, and at fixed concentration, $c$, and dilution, $d$, values corresponding to the two desired overdensity choices. Except for some curves in the left panels of Figure \ref{Fig - temperature profiles}, those with gas cores, $\varepsilon=0$, inside cuspy, $\alpha =1, \alpha=1.5$, dark matter halos, all other temperature profiles in Figure \ref{Fig - temperature profiles} exhibit a characteristic temperature peak within $0.03r_\mathrm{vir} - 0.1r_\mathrm{vir}$. This general shape is consistent with cluster temperature fits devised in X-ray observations \autocite[e.g.][]{Vikhlinin2006, Sun2009, Ghirardini2019, Lyskova2023}.

\subsubsection{\textcolor{black}{
The pressure profile of the ideal baryonic cluster halos}}

\textcolor{black}{
Similarly, by taking the density profiles of the ideal baryonic cluster halos into Equation \eqref{scale-free pressure}, the scale-free, equilibrium gas pressure profiles of these halos take the form:
\begin{equation}\label{ideal physical baryonic halo pressure}
	\frac{p(s, c, \alpha, \eta, d, \varepsilon)}{p_\mathrm{vir}} = \eta \mathcal{U}(c, \alpha, d, \varepsilon) \cdot \mathcal{I}(s, c, \alpha, \eta, d, \varepsilon).
\end{equation}}
\textcolor{black}{
To illustrate the predictive power of these simple analytic profiles, in Figure \ref{Fig - observational comparisons} we trace the total variation of these scale-free temperature and gas pressure profiles, from Equation \eqref{ideal physical baryonic halo temperature} and \eqref{ideal physical baryonic halo pressure}, respectively, over the physical parameter space corresponding to $\Delta=500$. For observational comparison, in the top panel of Figure \ref{Fig - observational comparisons} we compare our temperature profiles to the best-fit profile of observed X-COP galaxy clusters derived in \cite{Ghirardini2019}:— split into cool core clusters, the blue-dotted line, and non-cool core clusters, the orange dash-dotted line, and with their profiles re-scaled to meet our definition of $T_{500}$. Similarly, in the bottom panel of Figure \ref{Fig - observational comparisons}, we compare our gas pressure profiles to the universal profile from \cite{Arnaud2010}, shown by the purple dotted line, again re-scaled to meet out definition of $p_{500}$. }

\textcolor{black}{
In the panels in Figure \ref{Fig - observational comparisons}, the predicted variation within our model does a reasonable job at encompassing the observed fits, particularly within halocentric radii of $r\lesssim 0.5 r_{500}$. Toward larger halocentric radii, as seen in the top panel of Figure \ref{Fig - observational comparisons}, the temperature predictions in our model begin to deviate from the observed fits; in particular, our model systematically over-predicts the gas' temperature. To a lesser extent this trend is visible in the pressure comparison in the bottom panel of Figure \ref{Fig - observational comparisons}, where our model loses centring of the universal pressure profile. 
This over-prediction of the gas' temperature and pressure toward large halocentric radii is theoretically expected, due to non-thermal pressure contributors, such as turbulence and shocks, increasing toward the outer radii of real clusters, as studied in non-radiative hydrodynamic simulations \autocite[e.g.][]{Nelson2014, Angelinelli2020} and inferred observationally from measuring the hydrostatic bias of cluster masses \autocite[e.g.][]{Eckert2019, Sayers2021}. As these non-thermal pressure contributions were neglected in our model, our solutions for the gas' temperature and pressure will continue to satisfy the balance of hydrostatic equilibrium without their consideration, and so will expectedly overestimate the trends in real systems wherever non-thermal pressure becomes significant, as at these large halocentric radii. In future work, we will endeavour to improve our model by considering the non-thermal pressure profile of galaxy clusters.
}

\subsubsection{Convectional instabilities in the parameter space}

The minority of temperature profiles in Figure \ref{Fig - temperature profiles} that do not exhibit a temperature peak, as in the left panels, corresponding to $\varepsilon=0$, gas cores, instead show a flattening or increase of the temperature with decreasing halocentric radii. In particular, for cuspy, $\alpha = 1.5$ halos, shown in magenta, the temperature in the central region becomes divergent. To discuss the physical nature of these halos, we can quantify the effective polytropic index, $\Gamma$, for each halo over the parameter space, with this index defined by the logarithmic derivative:
\begin{equation}\label{polytropic index definition}
    \Gamma \equiv 1 + \frac{\mathrm{d} \ln T}{\mathrm{d} \ln \rho_\mathrm{gas}}.
\end{equation}
Physically, the index $\Gamma$ must not exceed the gas' adiabatic gas constant, $\gamma = 5/3$ for a monatomic gas, or the system will be unstable against convection. \textcolor{black}{This allows us to categorise whether a particular structural configuration of a cluster halo represents a stable configuration in its gas and dark matter components, or whether the halo will be unable to maintain such a configuration over time.}

When calculating the index $\Gamma$ over the parameter space of ideal baryonic cluster halos, we find that $\Gamma > 5/3$ for all clusters with a halo inner slope $\alpha=1.5$ and a gas core shallower than $\varepsilon \lesssim 0.05$; thus, the divergent temperature curves in Figure \ref{Fig - temperature profiles} are unstable halo configurations, as expected. When measuring $\Gamma$ down to very small halocentric radii, we find that $\Gamma > 5/3$ for all $\varepsilon = 0$ gas cores inside halos cuspier than $\alpha \gtrsim 1.04$. This implies that modelling a gas core inside a cuspier-than-NFW halo is an unstable parameter choice, particularly in the inner region of a cluster. Interestingly, this instability is corrected whenever the gas' inner slope becomes just slightly positive, i.e. $\varepsilon = 0^+$, for most of these halos, up to a maximum requirement of $\varepsilon \gtrsim 0.05$ for $\alpha=1.5$ halos, as already discussed. \textcolor{black}{As this unstable parameter region essentially comprises a boundary of the parameter space chosen in Table \ref{Parameter table}, it is not mathematically necessary to remove this region for our analysis. As such, we will maintain the parameter choices in Table \ref{Parameter table}, but we note that this unstable region of halo configurations should be kept in mind for application of our model to real clusters. }


\textcolor{black}{
Aside from this unstable region, the effective polytropic index of all other halos in our model is a monotonically increasing function of halocentric radii in the region $0.01 r_\mathrm{vir} \leq r \leq r_\mathrm{vir}$, with all halos converging toward a value of $\Gamma \simeq 1.1-1.2$ near the virial radius, in both overdensity conventions. This general trend is broadly consistent with observational constraints on the effective polytropic index of the intracluster gas in X-ray emitting clusters \autocite[see, e.g.][]{Ghirardini2019b}.}

\subsubsection{The emission-weighted temperature of the \\ ideal baryonic cluster halos}

\begin{figure*}[h!]
    \centering
    \includegraphics[width=\textwidth]{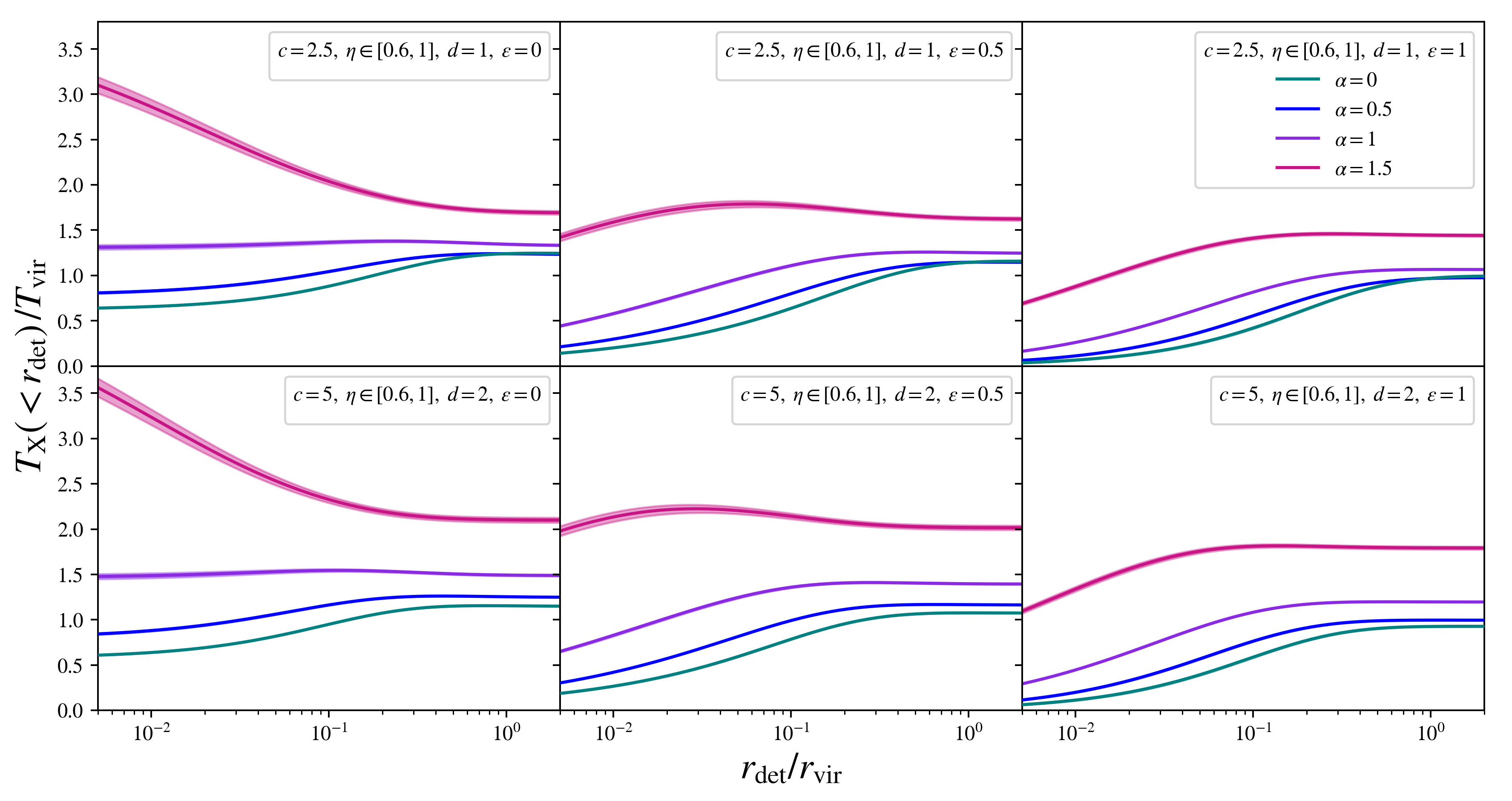}
    \caption{The emission-weighted temperature profiles for the ideal baryonic cluster halos, in scale-free form $\tau_1 \equiv T_\mathrm{X}(<r_\mathrm{det})/T_\mathrm{vir}$, traced over the scaled detection radius, $r_\mathrm{det}/r_\mathrm{vir}$. Each row varies the halo concentration, $c$, and the dilution, $d$, and each column varies the gas inner slope, $\varepsilon$. Within each box, each colour varies the halo inner slope, $\alpha$, with the solid coloured lines tracing a fraction of cosmological baryon content of $\eta = 0.8$, and the shaded colour region around each solid line (not visible for all curves) tracing this value continuously between $\eta=0.6$ and $\eta = 1$.}
    \label{Fig - TX profiles}
\end{figure*}

Taking the scale-free temperature profile for the ideal baryonic cluster halos, as in Equation \eqref{ideal physical baryonic halo temperature}, we can model the mean-weighted temperature observables for these halos over the desired parameter space. From Equation \eqref{scale-free X-ray emission-weighted temperature}, we predict the emission-weighted temperature profiles for these halos, in our scale-free framework, as:
\begin{equation}\label{ideal physical baryonic halo X-ray emission-weighted temperature}
	\frac{T_\mathrm{X}(<s_\mathrm{det}, c, \alpha, \eta, d, \varepsilon)}{T_\mathrm{vir}} = 3 \cdot \frac{\int _0 ^{s_\mathrm{det}} \frac{ \Bigl[\mathcal{I}(s, c, \alpha, \eta, d, \varepsilon) \Bigr]^{3/2}s ^{2} \mathrm{d}s}{\left\{s^{\varepsilon} \left[1 + \mathcal{C}(c, \alpha, d, \varepsilon)s\right]^{3-\varepsilon} \right\}^{1/2}}}{\int _0 ^{s_\mathrm{det}} \frac{ \Bigl[\mathcal{I}(s, c, \alpha, \eta, d, \varepsilon) \Bigr]^{1/2}s ^{2} \mathrm{d}s}{\left\{s^\varepsilon \left[1 + \mathcal{C}(c, \alpha, d, \varepsilon)s\right]^{3 - \varepsilon}\right\}^{3/2} }}.
\end{equation}
These scale-free temperature profiles, $T_\mathrm{X}(<r_\mathrm{det})/T_\mathrm{vir}$, are shown in Figure \ref{Fig - TX profiles}, as a function of the dimensionless detection radius, $s_\mathrm{det} \equiv r_\mathrm{det}/r_\mathrm{vir}$. 

\subsubsection{The mean gas mass-weighted temperature of the \\ ideal baryonic cluster halos}

\begin{figure*}[h!]
    \centering
    \includegraphics[width=\textwidth]{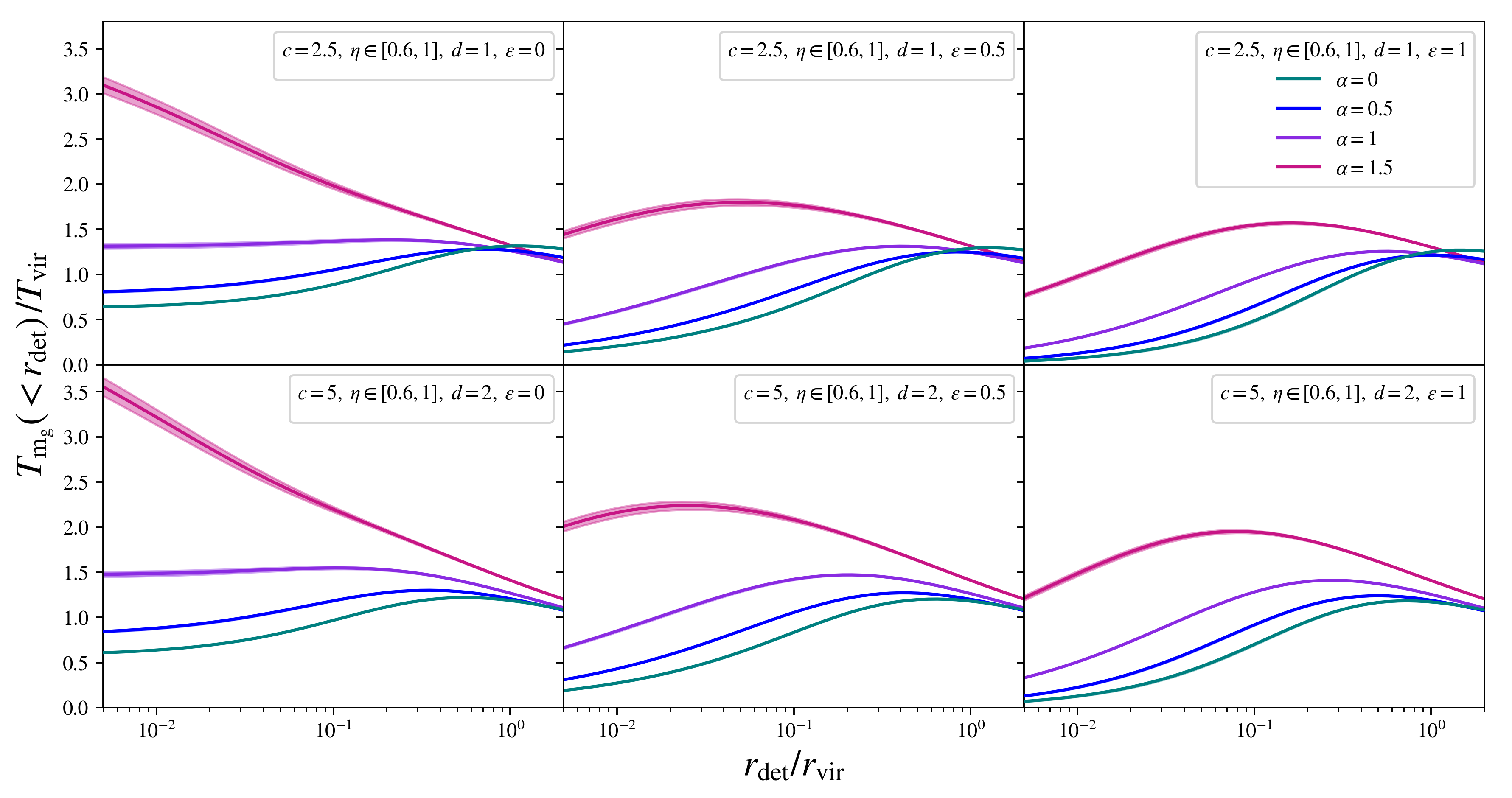}
    \caption{The mean gas mass-weighted temperature profiles for the ideal baryonic cluster halos, in scale-free form $\tau_2 \equiv T_\mathrm{m_g}(<r_\mathrm{det})/T_\mathrm{vir}$, traced over the scaled detection radius, $r_\mathrm{det}/r_\mathrm{vir}$. Each row varies the halo concentration, $c$, and the dilution, $d$, and each column varies the gas inner slope, $\varepsilon$. Within each box, each colour varies the halo inner slope, $\alpha$, with the solid coloured lines tracing a fraction of cosmological baryon content of $\eta = 0.8$, and the shaded colour region around each solid line (not visible for all curves) tracing this value continuously between $\eta=0.6$ and $\eta = 1$.}
    \label{Fig - Tmg profiles}
\end{figure*}

From Equation \eqref{mean gas mass-weighted temperature}, we predict the mean gas mass-weighted temperature profiles for the ideal baryonic cluster halos, in our scale-free framework, as:
\begin{equation}\label{ideal physical baryonic halo mean gas mass-weighted temperature}
	\frac{T_\mathrm{m_g}(<s_\mathrm{det}, c, \alpha, \eta, d, \varepsilon)}{T_\mathrm{vir}} = 3 \cdot \frac{\int _0 ^{s_\mathrm{det}}  \mathcal{I}(s, c, \alpha, \eta, d, \varepsilon) \, s ^{2} \mathrm{d}s}{\int _0^{s_\mathrm{det}} \frac{s^{2 - \varepsilon} \mathrm{d}s }{\left[1 + \mathcal{C}(c, \alpha, d, \varepsilon)s\right]^{3 - \varepsilon} } }.
\end{equation}
These scale-free temperature profiles, $T_\mathrm{m_g}(<r_\mathrm{det})/T_\mathrm{vir}$, are shown in Figure \ref{Fig - Tmg profiles}, as a function of the dimensionless detection radius, $s_\mathrm{det} \equiv r_\mathrm{det}/r_\mathrm{vir}$. 

\begin{figure*}[h!]
    \centering
    \includegraphics[width=\textwidth]{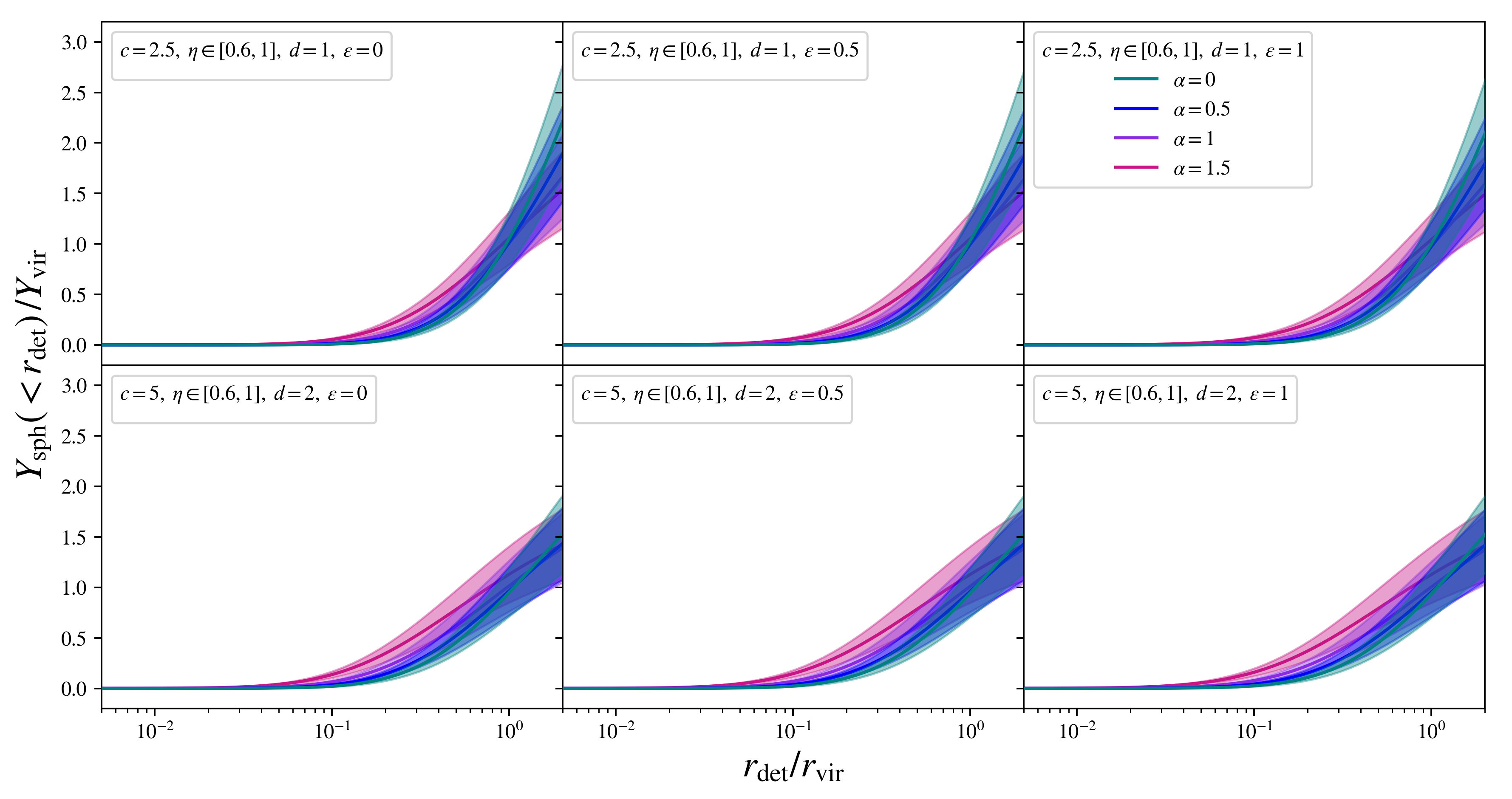}
    \caption{The spherically-integrated Sunyaev-Zeldovich signal for the ideal baryonic cluster halos, in scale-free form $\zeta_1 \equiv Y_\mathrm{sph}(<r_\mathrm{det})/Y_\mathrm{vir}$, traced over the scaled detection radius, $r_\mathrm{det}/r_\mathrm{vir}$. Each row varies the halo concentration, $c$, and the dilution, $d$, and each column varies the gas inner slope, $\varepsilon$. Within each box, each colour varies the halo inner slope, $\alpha$, with the solid coloured lines tracing a fraction of cosmological baryon content of $\eta = 0.8$, and the shaded colour region around each solid line tracing this value continuously between $\eta=0.6$ and $\eta = 1$.}
    \label{Fig - Ysph profiles}
\end{figure*}

\begin{figure*}[h!]
    \centering
    \includegraphics[width=\textwidth]{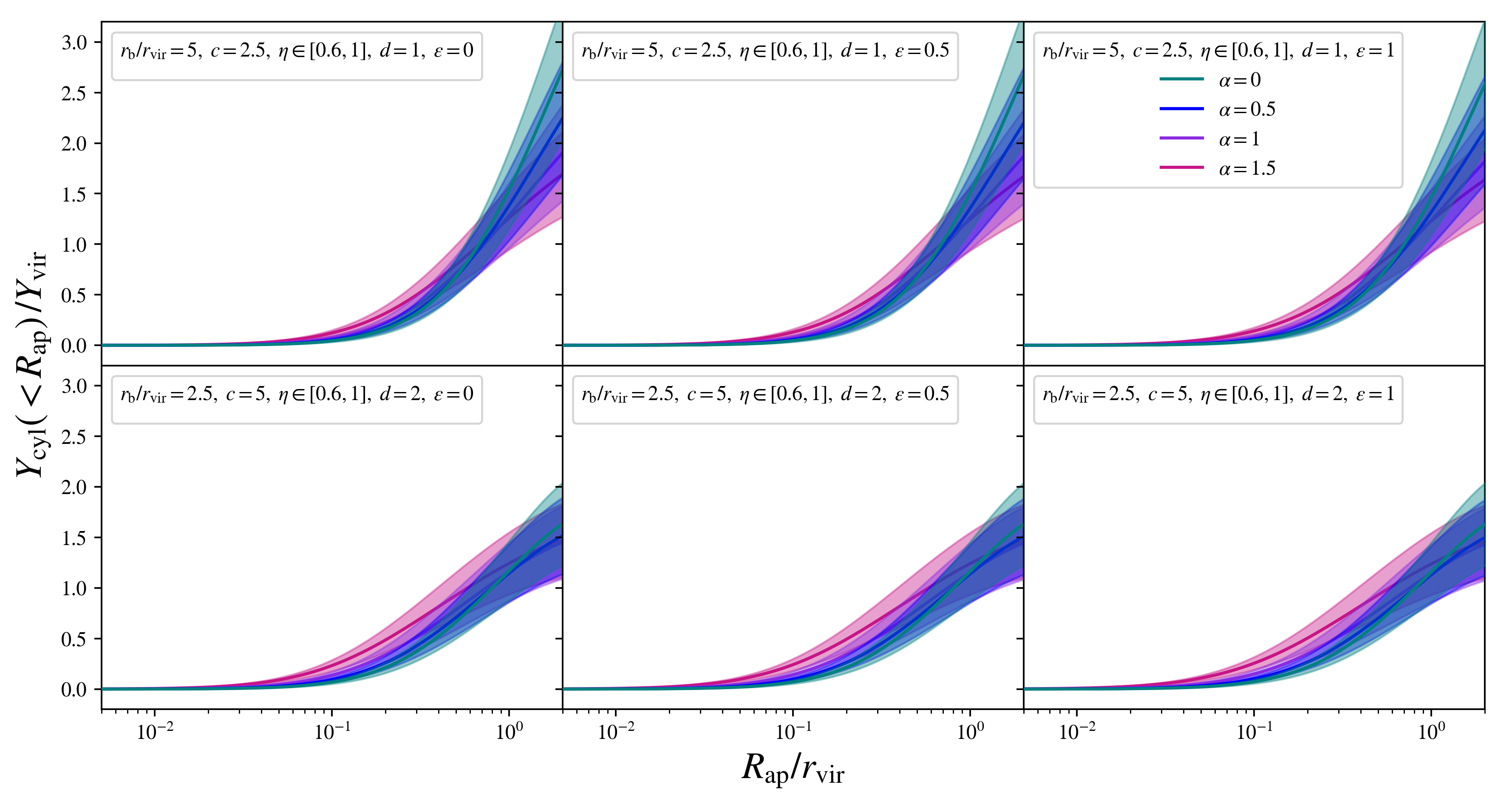}
    \caption{The cylindrically-integrated Sunyaev-Zeldovich signal for the ideal baryonic cluster halos, in scale-free form $\zeta_2 \equiv Y_\mathrm{cyl}(<R_\mathrm{ap})/Y_\mathrm{vir}$, traced over the scaled aperture radius, $R_\mathrm{ap}/r_\mathrm{vir}$. Each row varies the cluster boundary, in scale-free form $r_\mathrm{b}/r_\mathrm{vir}$, the halo concentration, $c$, and the dilution, $d$, and each column varies the gas inner slope, $\varepsilon$. Within each box, each colour varies the halo inner slope, $\alpha$, with the solid coloured lines tracing a fraction of cosmological baryon content of $\eta = 0.8$, and the shaded colour region around each solid line tracing this value continuously between $\eta=0.6$ and $\eta = 1$.}
    \label{Fig - Ycyl profiles}
\end{figure*}

\begin{figure*}[h!]
    \centering
    \includegraphics[width=\textwidth]{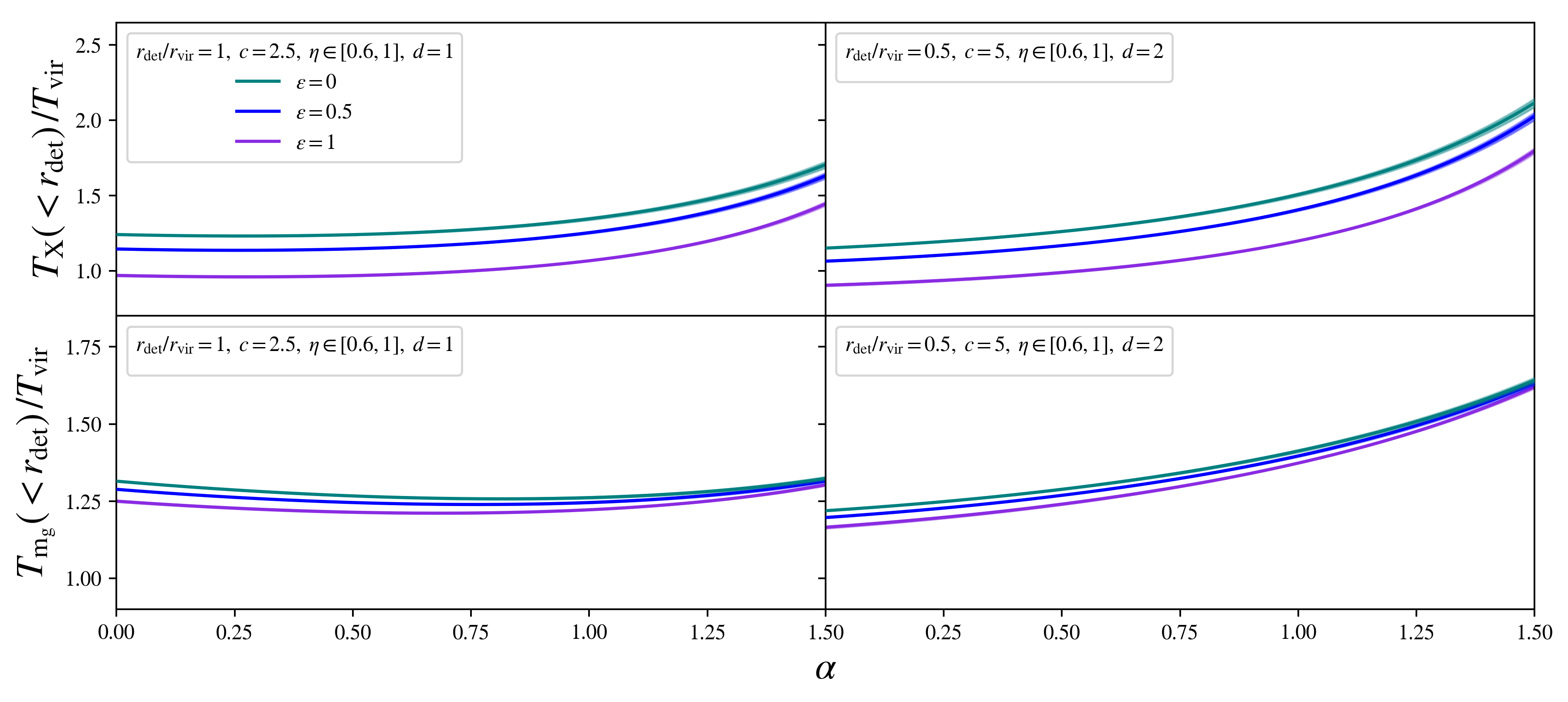}
    \caption{The emission-weighted and mean gas mass-weighted temperature profiles, in scale-free form $\tau_1  \equiv T_\mathrm{X}(<r_\mathrm{det})/T_\mathrm{vir}$ and $\tau_2 \equiv T_\mathrm{m_g}(<r_\mathrm{det})/T_\mathrm{vir}$, shown in the top and bottom panels, respectively, for the ideal baryonic cluster halos, evaluated at fixed detection radii, and traced over halo inner slopes, $\alpha$. Each column fixes the values of the detection radius, in scale-free form $r_\mathrm{det}/r_\mathrm{vir}$, the halo concentration, $c$, and the dilution, $d$, with these choices corresponding to a particular choice in the overdensity: the left panels for $\Delta=500$, and the right panels for $\Delta=200$. Within each box, each colour varies the gas inner slope, $\varepsilon$, with the solid coloured lines tracing a fraction of cosmological baryon content of $\eta = 0.8$, and the shaded colour region around each solid line (not visible for all curves) tracing this value continuously between $\eta=0.6$ and $\eta = 1$.}
    \label{Fig - weighted X-ray temperature profiles}
\end{figure*}

\subsubsection{The spherically-integrated Compton parameter of the \\ ideal baryonic cluster halos}

By integrating the product of the scale-free gas density and temperature profiles for the ideal baryonic cluster halos, as in Equations \eqref{ideal physical baryonic halo profile - baryonic gas} and \eqref{ideal physical baryonic halo temperature}, the integrated Compton parameters for these halos can be devised. From Equation \eqref{scale-free spherical SZ signal}, the spherically-integrated Compton parameter profiles for these halos are given by:

\begin{equation}\label{ideal physical baryonic halo SZ spherical signal}
\begin{aligned}
	\frac{Y_\mathrm{sph}(<s_\mathrm{det}, c, \alpha, \eta, d, \varepsilon)}{Y_\mathrm{vir}} &= 3 \eta \, \mathcal{U}(c, \alpha, d, \varepsilon) \\
    & \hspace{-10mm} \times \int _0 ^{s_\mathrm{det}} \mathcal{I}(s, c, \alpha, \eta, d, \varepsilon) s^2 \mathrm{d} s.
\end{aligned}
\end{equation}
These scale-free SZ profiles, $\zeta_1 \equiv Y_\mathrm{sph} (<r_\mathrm{det})/Y_\mathrm{vir}$, are shown in Figure \ref{Fig - Ysph profiles}, as a function of the dimensionless detection radius, $s_\mathrm{det} \equiv r_\mathrm{det}/r_\mathrm{vir}$. 

\subsubsection{The cylindrically-integrated Compton parameter of the \\ ideal baryonic cluster halos}

Finally, taking the profiles for the ideal baryonic cluster halos into Equation \eqref{scale-free cylindrical SZ signal}, the cylindrically-integrated Compton parameter profiles are given as:
\begin{equation}\label{ideal physical baryonic halo SZ cylindrical signal}
\begin{aligned}
	\frac{Y_\mathrm{cyl}(<S_\mathrm{ap}, s_\mathrm{b}, c, \alpha, \eta, d, \varepsilon)}{Y_\mathrm{vir}} &= 3 \eta \, \mathcal{U}(c, \alpha, d, \varepsilon) \,\,\, \times  \\
    & \hspace{-37mm} \Biggl[ \int _0 ^{s_\mathrm{b}} \mathcal{I}(s, c, \alpha, \eta, d, \varepsilon) s^2 \mathrm{d} s - \int _{S_\mathrm{ap}} ^{s_\mathrm{b}} \mathcal{I}(s, c, \alpha, \eta, d, \varepsilon) \sqrt{s^2 - S^2_\mathrm{ap}} s \mathrm{d} s\Biggr].
\end{aligned}
\end{equation}
These scale-free SZ profiles, $\zeta_2 \equiv Y_\mathrm{cyl} (<R_\mathrm{ap})/Y_\mathrm{vir}$, are shown in Figure \ref{Fig - Ycyl profiles}, as a function of the dimensionless aperture radius, $S_\mathrm{ap} \equiv R_\mathrm{ap}/r_\mathrm{vir}$. 

\subsection{3.2. \quad Constraints on $\tau_1$, $\tau_2$, $\zeta_1$ and $\zeta_2$}

\begin{table*}
\begin{tabular}{ |p{2.8cm}|p{2cm}|p{1.7cm}|p{1cm}||p{1.7cm}|p{2cm}| }
\hline
Overdensity parameter & Detection radius & Concentration & Dilution & \multicolumn{2}{|c|}{ Bounds in $\tau_1\equiv T_\mathrm{X}(<r_\mathrm{det})/T_\mathrm{vir}$ } \\ [1ex]
\hline\hline
\rowcolor{white}  $\Delta=200$ & $r_\mathrm{det} = 0.5r_{200}$ & $c=5$ & $d=2$ &  $\tau_\mathrm{1, min} = 0.899$ & $\tau_\mathrm{1, max} = 2.145$ \\ [2ex]
\hline
\rowcolor{white} $\Delta=500$ & $r_\mathrm{det} = r_{500}$ & $c=2.5$ & $d=1$ & $\tau_\mathrm{1, min} = 0.956$ & $\tau_\mathrm{1, max} =1.725$ \\ [2ex]
\hline
\end{tabular}

\vspace{0.3cm}

\begin{tabular}{ |p{2.8cm}|p{2cm}|p{1.7cm}|p{1cm}||p{1.7cm}|p{2cm}| }
\hline
Overdensity parameter & Detection radius & Concentration & Dilution & \multicolumn{2}{|c|}{ Bounds in $\tau_2\equiv T_\mathrm{m_g}(<r_\mathrm{det})/T_\mathrm{vir}$ } \\ [1ex]
\hline\hline
\rowcolor{white}  $\Delta=200$ & $r_\mathrm{det} = 0.5r_\mathrm{200}$ & $c=5$ & $d=2$ &  $\tau_\mathrm{2, min} = 1.162$ & $\tau_\mathrm{2, max} =1.650$ \\ [2ex]
\hline
\rowcolor{white} $\Delta=500$ & $r_\mathrm{det} = r_\mathrm{500}$ & $c=2.5$ & $d=1$ & $\tau_\mathrm{2, min} = 1.210$ & $\tau_\mathrm{2, max} =1.329$ \\ [2ex]
\hline
\end{tabular}
\caption{The constraints placed on $\tau_1 \equiv T_\mathrm{X}(<r_\mathrm{det})/T_\mathrm{vir}$ and $\tau_2 \equiv T_\mathrm{m_g}(<r_\mathrm{det})/T_\mathrm{vir}$ over the parameter space of the ideal baryonic cluster halos, each evaluated at two conventions in the overdensity, $\Delta=200$ and $\Delta=500$.}

\label{Table - tau bounds}
\end{table*}

\begin{table*}[h!]
\begin{tabular}{ |p{2.8cm}|p{2cm}|p{1.7cm}|p{1cm}||p{1.7cm}|p{2cm}| }
\hline
Overdensity parameter & Detection radius & Concentration & Dilution & \multicolumn{2}{|c|}{ Bounds in $\zeta_1\equiv Y_\mathrm{sph}(<r_\mathrm{det})/Y_\mathrm{vir}$ } \\ [1ex]
\hline\hline
\rowcolor{white}  $\Delta=200$ & $r_\mathrm{det} = 0.5r_{200}$ & $c=5$ & $d=2$ &  $\zeta_\mathrm{1, min} = 0.336$ & $\zeta_\mathrm{1, max} = 0.983$ \\ [2ex]
\hline
\rowcolor{white} $\Delta=500$ & $r_\mathrm{det} = r_{500}$ & $c=2.5$ & $d=1$ & $\zeta_\mathrm{1, min} = 0.726$ & $\zeta_\mathrm{1, max} = 1.319$ \\ [2ex]
\hline
\end{tabular}

\vspace{0.3cm}

\begin{tabular}{ |p{2.8cm}|p{2cm}|p{2.1cm}|p{1.7cm}|p{1cm}||p{1.7cm}|p{2cm}| }
\hline
Overdensity parameter & Aperture radius & Cluster boundary & Concentration & Dilution & \multicolumn{2}{|c|}{ Bounds in $\zeta_2\equiv Y_\mathrm{cyl}(<R_\mathrm{ap})/Y_\mathrm{vir}$ } \\ [1ex]
\hline\hline
\rowcolor{white}  $\Delta=200$ & $R_\mathrm{ap} = 0.5r_\mathrm{200}$ & $r_\mathrm{b} = 2.5 r_{200}$ & $c=5$ & $d=2$ &  $\zeta_\mathrm{2, min} = 0.492$ & $\zeta_\mathrm{2, max} = 1.161$ \\ [2ex]
\hline
\rowcolor{white} $\Delta=500$ & $R_\mathrm{ap} = r_\mathrm{500}$ & $r_\mathrm{b} = 5 r_{500}$ & $c=2.5$ & $d=1$ & $\zeta_\mathrm{2, min} = 0.916$ & $\zeta_\mathrm{2, max} = 1.928$ \\ [2ex]
\hline
\end{tabular}
\caption{The constraints placed on $\zeta_1 \equiv Y_\mathrm{sph}(<r_\mathrm{det})/Y_\mathrm{vir}$ and $\zeta_2 \equiv Y_\mathrm{cyl}(<R_\mathrm{ap})/Y_\mathrm{vir}$ over the parameter space of the ideal baryonic cluster halos, each evaluated at two conventions in the overdensity, $\Delta=200$ and $\Delta=500$.}
\label{Table - zeta bounds}
\end{table*}

With these scale-free predictions for the X-ray and SZ observables, the dimensionless parameters $\tau_1$, $\tau_2$, $\zeta_1$ and $\zeta_2$ are constrained inside minimum and maximum values, when evaluating these profiles in each of the desired regimes. 

\subsubsection{Constraints on $\tau_1$ and $\tau_2$ within each regime}

The two regimes in which the dimensionless weighted temperature parameters, $\tau_1$ and $\tau_2$, are to be bounded can be traced within Figures \ref{Fig - TX profiles} and \ref{Fig - Tmg profiles}, respectively. Each row of panels in each of these figures fixes the halo concentration, $c$, and dilution, $d$, to the set of values required in each regime, such that when these profiles are evaluated at a chosen detection radius, in scale-free form $r_\mathrm{det}/r_\mathrm{vir}$, the bounds in $\tau_1$ and $\tau_2$ will be determined. Figure \ref{Fig - weighted X-ray temperature profiles} evaluates these scale-free weighted temperatures from Figures \ref{Fig - TX profiles} and \ref{Fig - Tmg profiles} as a continuous function of the halo inner slope, $\alpha$, at detection radii corresponding to the two regimes --- $r_\mathrm{det}/r_\mathrm{vir} = 0.5$ for $\Delta=200$ and $r_\mathrm{det}/r_\mathrm{vir} =1$ for $\Delta=500$ --- producing four distinct windows, two for each observable. In the panels of Figure \ref{Fig - weighted X-ray temperature profiles}, within each box, the scale-free weighted temperatures are always bounded between the minimum, $\varepsilon = 1$, in purple, and the maximum, $\varepsilon = 0$, in teal, fixed gas slope curves. 
As such, the values $\tau_\mathrm{1, min}$ and $\tau_\mathrm{1, max}$, or $\tau_\mathrm{2, min}$ and $\tau_\mathrm{2, max}$, are simply the minimum and maximum values of the parameter space traced between these curves, in each box, corresponding to each of the two regimes, for each parameter. These results produce the constraints in $\tau_1$ and $\tau_2$ as summarised in Table \ref{Table - tau bounds}.

\subsubsection{Constraints on $\zeta_1$ and $\zeta_2$ within each regime}

Similarly, the two regimes in which the dimensionless SZ parameters $\zeta_1$ and $\zeta_2$ are to be bounded can be traced within Figures \ref{Fig - Ysph profiles} and \ref{Fig - Ycyl profiles}, respectively. As before, each row of panels fixes the halo concentration, $c$, and dilution, $d$, to the set of values required in each of the required regimes. At the desired detection and aperture radii --- $r_\mathrm{det}/r_\mathrm{vir} = 0.5$ and $R_\mathrm{ap}/r_\mathrm{vir} = 0.5$ when $\Delta=200$ and $r_\mathrm{det}/r_\mathrm{vir} = 1$ and $R_\mathrm{ap}/r_\mathrm{vir} = 1$ when $\Delta=500$ --- the integrated Compton parameter profiles in Figures \ref{Fig - Ysph profiles} and \ref{Fig - Ycyl profiles} are considerably degenerate for the parameters shown across these panels. By taking the minimum and maximum values of $\zeta_1 \equiv Y_\mathrm{sph}(<r_\mathrm{det})/Y_\mathrm{vir}$ and $Y_\mathrm{cyl}(<R_\mathrm{ap})/Y_\mathrm{vir}$ at each desired detection and aperture radius, respectively, the bounds summarised in Table \ref{Table - zeta bounds} are calculated. 

\section{4. \quad Results}
\vspace{2mm}
\subsection{4.1. \quad The $M_\mathrm{vir} - T_\mathrm{X}$ and $M_\mathrm{vir} - T_\mathrm{m_g}$ scaling relations}

\begin{figure*}[h!]
    \centering
    \includegraphics[width=\textwidth]{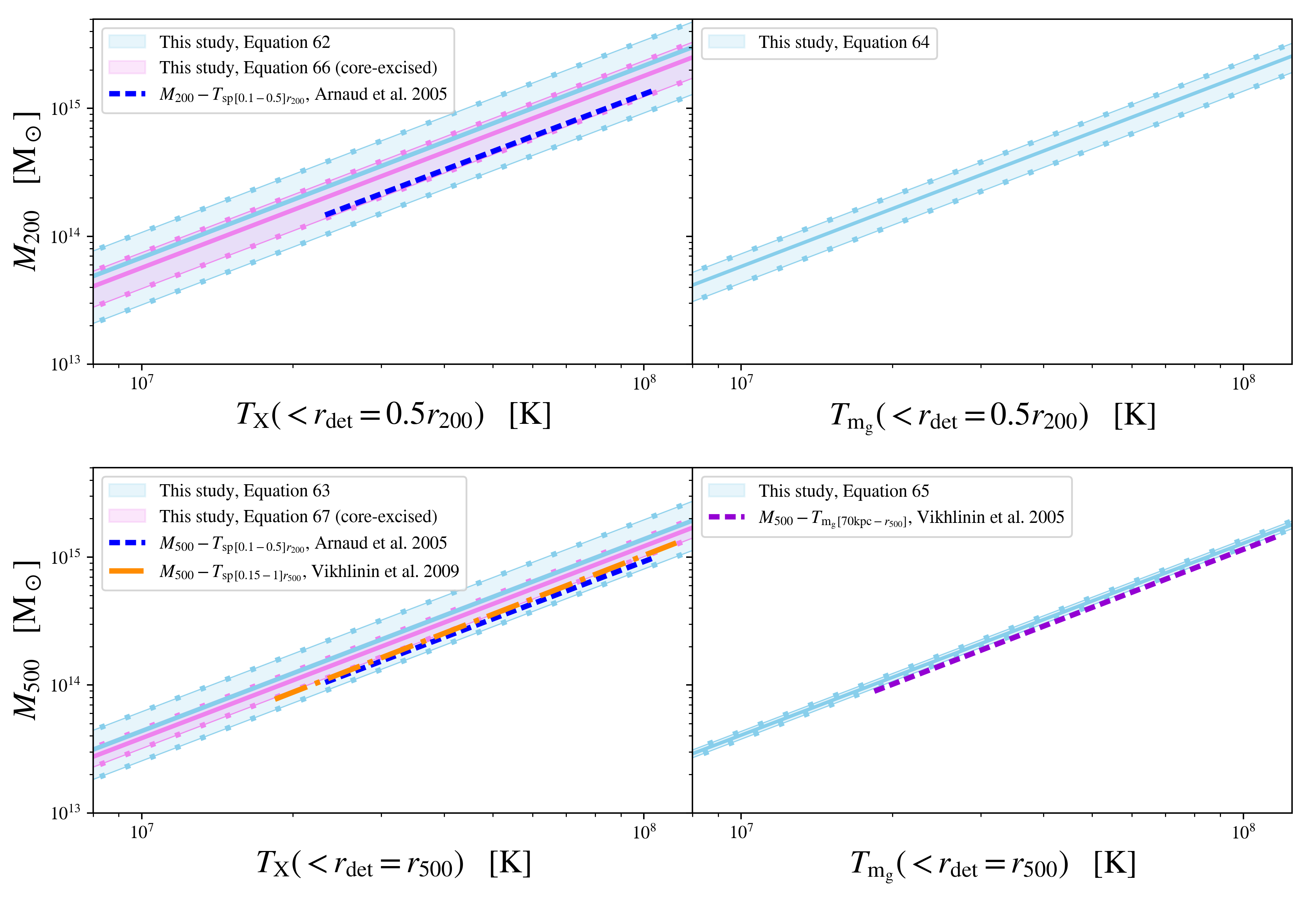}
    \caption{Our predictions for the halo mass - temperature scaling relations, in terms the emission-weighted temperature, $M_{200} - T_\mathrm{X}$ and $M_{500} - T_\mathrm{X}$, in the left panels, and the mean gas mass-weighted temperature, $M_{200} - T_\mathrm{m_g}$ and $M_{500} - T_\mathrm{m_g}$, in the right panels. The light blue intervals trace the uncertainties in each of these scaling relations, as quantified by constraints in the dimensionless parameters, $\tau_1$ and $\tau_2$, given in Table \ref{Table - tau bounds}, derived over specified halo parameters. In the left panels, the additional pink interval traces the scaling relations formed by excising the cluster's central core, given by the constraints in Table \ref{Table - tau core excised}. For each of these coloured intervals, the solid dotted lines enclosing each interval correspond to the minimum and maximum bounds in the scaling relation, with the solid central line tracing its mid-range value. These predictions are shown in comparison to observational fits to these scaling relations, from \cite{Arnaud2005} (the blue dotted lines), \cite{Vikhlinin2006} (the orange dash-dotted line) and \cite{Vikhlinin2009} (the purple dotted line). }
    \label{Fig - virial mass scaling relations}
\end{figure*}

From our analysis, we are now able to present the predicted scaling relations of the halo mass with its weighted temperatures observables. In particular, from the minimum and maximum values of $\tau_1$ and $\tau_2$ detailed in Table \ref{Table - tau bounds}, the scaling relations of the halo mass with the weighted temperatures, $T_\mathrm{X}$ and $T_\mathrm{m_g}$, as devised in Section 2.4 --- Equations \eqref{M200 TX mass bound} and \eqref{M500 TX mass bound}  --- will be constrained. 
These predictions are illustrated in Figure \ref{Fig - virial mass scaling relations}, 
with the left panels showing the $M_{200} - T_\mathrm{X}$ and $M_{500} - T_\mathrm{X}$ scaling relations, and the right panels showing the  $M_{200} - T_\mathrm{m_g}$ and $M_{500} - T_\mathrm{m_g}$ scaling relations. Each of these scaling relations and their predicted uncertainties are presented analytically below. 


\subsubsection{The $M_{200} - T_\mathrm{X}$ and $M_{500} - T_\mathrm{X}$ scaling relations}

In terms of the mid-range value and the associated uncertainty, the $M_{200} - T_\mathrm{X}$ scaling relation is predicted as:
\begin{equation}\label{M200 scaling relation - emission-weighted}
\begin{aligned}
    M_{200} &= \left(6.81 \pm 3.90\right) \cdot  \left[\frac{T_\mathrm{X}}{10^7\,\mathrm{K}}\right]^{3/2} \cdot 10^{13} \, \mathrm{M}_\odot,
\end{aligned}
\end{equation}
and similarly, the $M_{500} - T_\mathrm{X}$ scaling relation, as:
\begin{equation}\label{M500 scaling relation - emission-weighted}
\begin{aligned}
    M_{500} &= \left(4.36 \pm 1.81\right) \cdot  \left[\frac{T_\mathrm{X}}{10^7\,\mathrm{K}}\right]^{3/2} \cdot 10^{13} \, \mathrm{M}_\odot,
\end{aligned}
\end{equation}
with each of these relations attaining an uncertainty of $57.3\%$ and $41.6\%$, respectively. 

\subsubsection{The $M_{200} - T_\mathrm{m_g}$ and $M_{500} - T_\mathrm{m_g}$ scaling relations}

Similarly, the $M_{200} - T_\mathrm{m_g}$ and $M_{500} - T_\mathrm{m_g}$ scaling relations are predicted as:
\begin{equation}\label{M200 scaling relation - mean gas mass-weighted}
\begin{aligned}
    M_{200} &= \left(5.80 \pm 1.49\right) \cdot  \left[\frac{T_\mathrm{m_g}}{10^7\,\mathrm{K}}\right]^{3/2} \cdot 10^{13} \, \mathrm{M}_\odot,
\end{aligned}
\end{equation}
and, as:
\begin{equation}\label{M500 scaling relation - mean gas mass-weighted}
\begin{aligned}
    M_{500} &= \left(4.05 \pm 0.28\right) \cdot  \left[\frac{T_\mathrm{m_g}}{10^7\,\mathrm{K}}\right]^{3/2} \cdot 10^{13} \, \mathrm{M}_\odot,
\end{aligned}
\end{equation}
with each attaining an uncertainty of $25.7\%$ and $7.0\%$, respectively. 

\subsubsection{Comparison of our predictions to observed X-ray relations}

\begin{table*}
\begin{tabular}{ |p{2.8cm}|p{2cm}|p{2cm}|p{1.7cm}|p{1cm}||p{1.7cm}|p{2cm}| }
\hline
Overdensity parameter & Detection radius & Excised core & Concentration & Dilution & \multicolumn{2}{|c|}{ Bounds in $\tau_1\equiv T_\mathrm{X}[r_\mathrm{ex}-r_\mathrm{det}]/T_\mathrm{vir}$ } \\ [1ex]
\hline\hline
\rowcolor{white}  $\Delta=200$ & $r_\mathrm{det} = 0.5r_{200}$ & $r_\mathrm{ex} = 0.1r_{200}$ & $c=5$ & $d=2$ &  $\tau_\mathrm{1, min} = 1.146$ & $\tau_\mathrm{1, max} = 1.766$ \\ [2ex]
\hline
\rowcolor{white} $\Delta=500$ & $r_\mathrm{det} = r_{500}$ & $r_\mathrm{ex} = 0.15r_{500}$ & $c=2.5$ & $d=1$ & $\tau_\mathrm{1, min} = 1.181$ & $\tau_\mathrm{1, max} = 1.487$ \\ [2ex]
\hline
\end{tabular}
\caption{The constraints placed on $\tau_1 \equiv T_\mathrm{X}[r_\mathrm{ex} - r_\mathrm{det}]/T_\mathrm{vir}$ over the parameter space of the ideal baryonic cluster halos, each evaluated at two conventions in the overdensity, $\Delta=200$ and $\Delta=500$, when excising a central core of halocentric radius $r_\mathrm{ex}$.}
\label{Table - tau core excised}
\end{table*}

\begin{figure*}[h!]
    \centering
    \includegraphics[width=\textwidth]{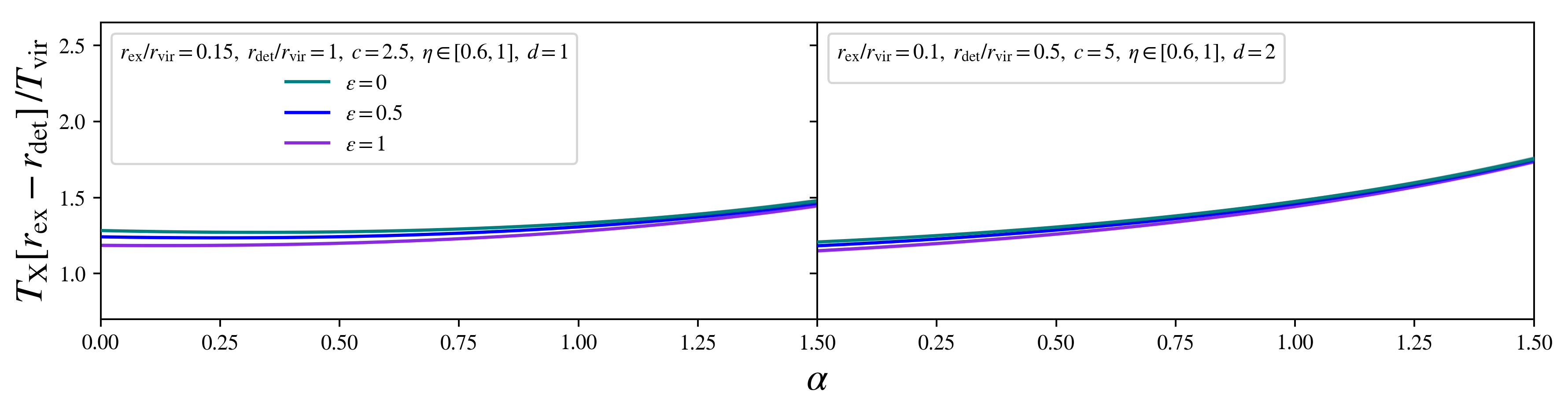}
    \caption{The core-excised emission-weighted temperature profiles, in scale-free form $\tau_1  \equiv T_\mathrm{X}[r_\mathrm{ex} - r_\mathrm{det}]/T_\mathrm{vir}$, for the ideal baryonic cluster halos, evaluated at fixed detection radii, and traced over halo inner slopes, $\alpha$. Each panel fixes the core-excised region, in scale-free form $r_\mathrm{ex}/r_\mathrm{vir}$, the detection radius, in scale-free form $r_\mathrm{det}/r_\mathrm{vir}$, the halo concentration, $c$, and the dilution, $d$, with these choices corresponding to a particular choice in the overdensity: the left panel for $\Delta=500$, and the right panel for $\Delta=200$. Within each box, each colour varies the gas inner slope, $\varepsilon$, with the solid coloured lines tracing a fraction of cosmological baryon content of $\eta = 0.8$, with the variation between $\eta=0.6$ and $\eta=1$ so weak that no shaded colour region around these solid lines is visible.}
    \label{Fig - core-excised temperature profiles}
\end{figure*}

\textcolor{black}{
Where possible, in Figure \ref{Fig - virial mass scaling relations}, our predicted scaling relations are compared to relations devised in the literature of X-ray cluster observations: from \cite{Arnaud2005}, \cite{Vikhlinin2006} and \cite{Vikhlinin2009}. Of note, some of these studies use the cluster's spectroscopic temperature, $T_\mathrm{sp}$, to fit the scaling relation. This temperature measure is usually assumed to be comparable to the emission-weighted temperature, $T_\mathrm{X}$, hence the comparison in the left panels of Figure \ref{Fig - virial mass scaling relations}; however, $T_\mathrm{sp}$ is known to underestimate $T_\mathrm{X}$ \autocite[see, e.g.][]{Rasia2005}, and so this comparison is not one-to-one. }

\textcolor{black}{
Despite the differences between our model's predictions and the observational comparisons in Figure \ref{Fig - virial mass scaling relations}, it remains clear that our predictions generally contain the expected trends, whilst systematically overestimating the cluster's halo mass. Only our $M_{500} - T_\mathrm{m_g}$ prediction, in the lower right panel, excludes the observed fit from \cite{Vikhlinin2006}, which lies $\sim 6\%$ below our lower bound. This small deviation is reasonable, and likely attributed to systematics involved when comparing the normalisation of our model and the observational measurement of $r_{500}$ for calculating $T_\mathrm{m_g}$. }

\textcolor{black}{
Moreover, whilst the observed scaling relations are all expected to underestimate the halo mass, due to the hydrostatic bias, our modelling is also expected underestimate the halo mass, by the same bias, as in our model we do not consider non-thermal pressure, which will decrease the temperature required to attain a given halo mass. As such, the hydrostatic bias itself cannot explain our systematic overestimates in the scaling relations; instead, this is attributed to the chosen parameter space and the generous range chosen in halo structures, driving a larger scatter, which is mitigated in the real universe as real clusters do not not uniformly sample this parameter space.}

\subsubsection{Excising the central core in the $M_{200} - T_\mathrm{X}$ and $M_{500} - T_\mathrm{X}$ \\ scaling relations}

\textcolor{black}{
In the observational comparisons shown in Figure \ref{Fig - virial mass scaling relations}, the cluster region in which the weighted temperature is measured differs for each study, all excising some central region, e.g. $0.1r_{500}$, $70$ $\mathrm{kpc}$, or some other excised halocentric radius, $r_\mathrm{ex}$. In our model, excising a central core has a significant impact on the $T_\mathrm{X}$ estimate, and thus the constraints on $\tau_1$, due to its dependence on integrating the temperature $T^{3/2}$, as in Equation \eqref{emission-weighted temperature}, which in the central region, for non-cool core clusters with high central temperatures, will have a non-negligible contribution to the value of $T_\mathrm{X}$ attained within the detection radius. For this reason, in Figure \ref{Fig - virial mass scaling relations}, we show an additional prediction for both the $M_{200} - T_\mathrm{X}$ and $M_{500} - T_\mathrm{X}$ scaling relations, shown in pink, that each excise a central region of $0.1r_{200}$ and $0.15r_{500}$, respectively. These scaling relations are constrained by new values of $\tau_\mathrm{1,min}$ and $\tau_\mathrm{1,max}$, given in Table \ref{Table - tau core excised}, as derived in analogue to our previous analysis, and traced by the scale-free profiles $\tau_1 \equiv T_\mathrm{X}[r_\mathrm{ex} - r_\mathrm{det}]/T_\mathrm{vir}$ that are shown in Figure \ref{Fig - core-excised temperature profiles}. These core-excised scaling relations take the form:
\begin{equation}\label{M200 scaling relation - emission-weighted, core excised}
\begin{aligned}
    M_{200} &= \left(5.67 \pm 1.78\right) \cdot  \left[\frac{T_\mathrm{X}}{10^7\,\mathrm{K}}\right]^{3/2} \cdot 10^{13} \, \mathrm{M}_\odot,
\end{aligned}
\end{equation}
and:
\begin{equation}\label{M500 scaling relation - emission-weighted, core excised}
\begin{aligned}
    M_{500} &= \left(3.84 \pm 0.66\right) \cdot  \left[\frac{T_\mathrm{X}}{10^7\,\mathrm{K}}\right]^{3/2} \cdot 10^{13} \, \mathrm{M}_\odot,
\end{aligned}
\end{equation}
each attaining a significantly lower uncertainty than the previous estimates, by approximately half, at $31.3\%$ and $17.1\%$, respectively. Thus, expectedly, by excluding the variance of temperature curves in the central region, these scaling relations provide a stronger halo mass constraint. Moreover, these core-excised predictions lower the mid-range value of each scaling relation, and so become closer to the observational fits, as expected in this more aligned comparison.}

\subsection{4.2. \quad The $M_\mathrm{vir} - Y_\mathrm{sph}$ and $M_\mathrm{vir} - Y_\mathrm{cyl}$ scaling relations}

\begin{figure*}[h!]
    \centering
    \includegraphics[width=\textwidth]{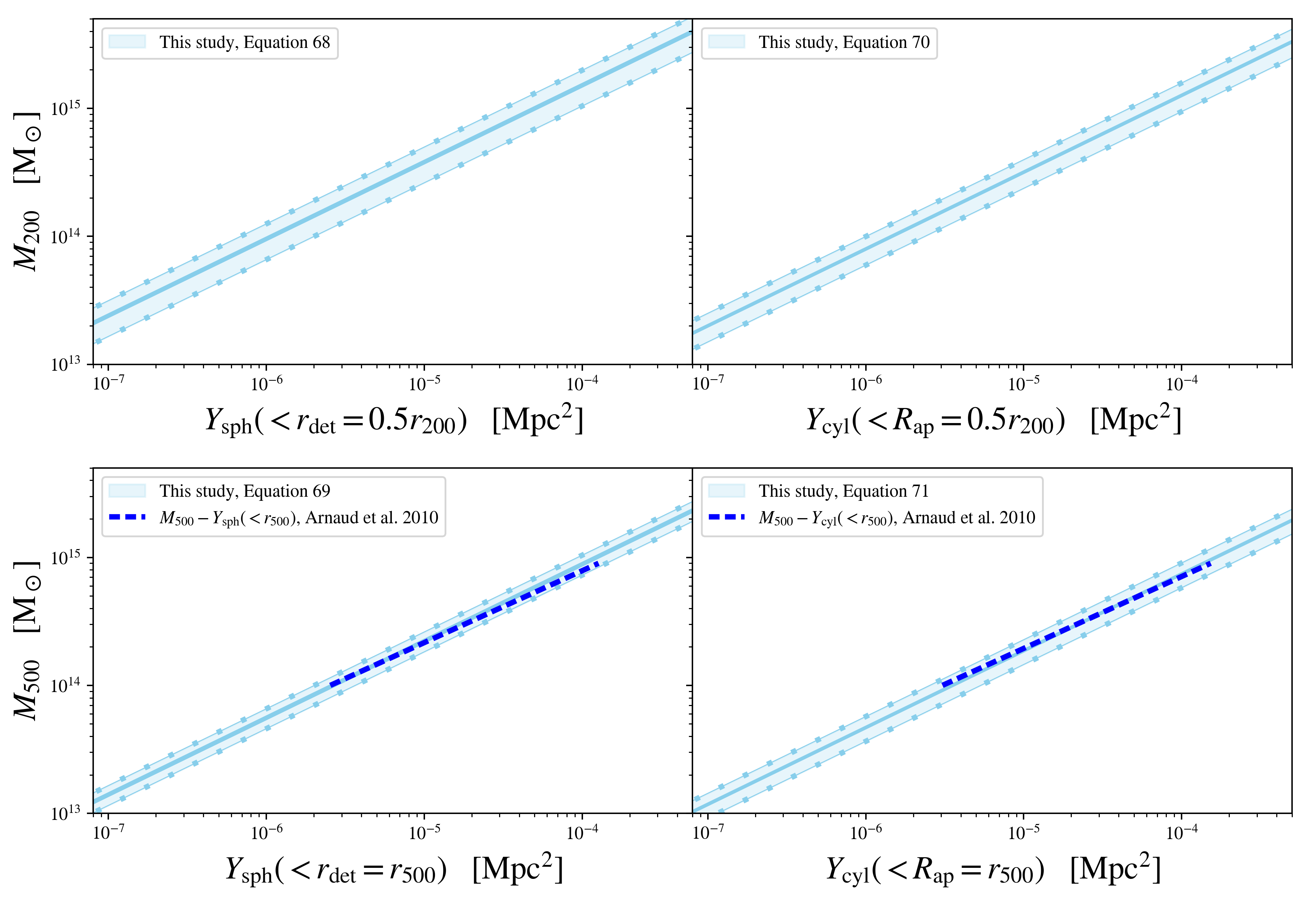}
    \caption{Our predictions for the halo mass - Sunyaev-Zeldovich scaling relations, in terms the spherically-integrated Compton parameter, $M_{200} - Y_\mathrm{sph}$ and $M_{500} - Y_\mathrm{sph}$, in the left panels, and the cylindrically-integrated Compton parameter, $M_{200} - Y_\mathrm{cyl}$ and $M_{500} - Y_\mathrm{cyl}$, in the right panels. The light blue intervals trace the uncertainties in each of these scaling relations, as quantified by constraints in the dimensionless parameters, $\zeta_1$ and $\zeta_2$, given in Table \ref{Table - zeta bounds}, derived over specified halo parameters. The solid dotted lines enclosing each interval correspond to the minimum and maximum bounds in the scaling relation, with the solid central line tracing its mid-range value. These predictions are shown in comparison to the scaling relations predicted by \cite{Arnaud2010} (the blue dotted lines).}
    \label{Fig - SZ virial mass scaling relations}
\end{figure*}

Furthermore, we now present the predicted scaling relations of the halo mass with the integrated SZ observables of galaxy clusters. In analogue, from the minimum and maximum values of $\zeta_1$ and $\zeta_2$ detailed in Table \ref{Table - zeta bounds}, the scaling relations of the halo mass with the integrated Compton parameters, $Y_\mathrm{sph}$ and $Y_\mathrm{cyl}$, as devised in Section 2.4 --- Equations \eqref{M200 SZ spherical mass bound} and \eqref{M500 SZ spherical mass bound}  --- will be constrained. In Figure \ref{Fig - SZ virial mass scaling relations}, these predictions are shown, 
with the scaling relations $M_{200} - Y_\mathrm{sph}$ and $M_{500} - Y_\mathrm{sph}$ shown in the left panels, 
and the scaling relations $M_{200} - Y_\mathrm{cyl}$ and $M_{500} - Y_\mathrm{cyl}$ shown in the right panels. Again, these scaling relations and their predicted uncertainties are given analytically, as detailed below. 

\subsubsection{The $M_{200} - Y_\mathrm{sph}$ and $M_{500} - Y_\mathrm{sph}$ scaling relations}

As before, taking the mid-range value and the associated uncertainty, the $M_{200} - Y_\mathrm{sph}$ scaling relation is predicted as:
\begin{equation}\label{M200 scaling relation - SZ, spherical}
\begin{aligned}
    M_{200} &= \left(9.54 \pm 2.97\right) \cdot  \left[\frac{Y_\mathrm{sph}}{10^{-6}\,\mathrm{Mpc^2}}\right]^{3/5} \cdot 10^{13} \, \mathrm{M}_\odot,
\end{aligned}
\end{equation}
and similarly, the $M_{500} - Y_\mathrm{sph}$, as:
\begin{equation}\label{M500 scaling relation - SZ, spherical}
\begin{aligned}
    M_{500} &= \left(5.58 \pm 0.99\right) \cdot  \left[\frac{Y_\mathrm{sph}}{10^{-6}\,\mathrm{Mpc^2}}\right]^{3/5} \cdot 10^{13} \, \mathrm{M}_\odot,
\end{aligned}
\end{equation}
with each attaining an uncertainty of $31.1\%$ and $17.7\%$, respectively.

\subsubsection{The $M_{200} - Y_\mathrm{cyl}$ and $M_{500} - Y_\mathrm{cyl}$ scaling relations}

Finally, in the same way, the predicted scaling relations $M_{200} - Y_\mathrm{cyl}$ and $M_{500} - Y_\mathrm{cyl}$ are given by:
\begin{equation}\label{M200 scaling relation - SZ, cylindrical}
\begin{aligned}
    M_{200} &= \left(7.96 \pm 2.01 \right) \cdot  \left[\frac{Y_\mathrm{cyl}}{10^{-6}\,\mathrm{Mpc^2}}\right]^{3/5} \cdot 10^{13} \, \mathrm{M}_\odot,
\end{aligned}
\end{equation}
and:
\begin{equation}\label{M500 scaling relation - SZ, cylindrical}
\begin{aligned}
    M_{500} &= \left(4.68 \pm 1.03\right) \cdot  \left[\frac{Y_\mathrm{cyl}}{10^{-6}\,\mathrm{Mpc^2}}\right]^{3/5} \cdot 10^{13} \, \mathrm{M}_\odot,
\end{aligned}
\end{equation}
with each of these scaling relations attaining an uncertainty of $25.2\%$ and $22.0\%$, respectively.

\subsubsection{Comparison of our predictions to observed SZ relations}

\textcolor{black}{
As shown in Figure \ref{Fig - SZ virial mass scaling relations}, our predicted scaling relations are compared to the SZ scaling relations devised in \cite{Arnaud2010}, for the scaling relation with the halo mass $M_{500}$. Unlike in the previous comparison, in this comparison the observed fits can be compared like-for-like with our predictions, as the integrated Compton parameters fit to each scaling are identical in definition to the observables chosen in our model. Here, our predicted intervals perfectly contain the observed fits, illustrating the strong predictive power of our simple model, and the self-similar expectation for this scaling relation. 
}

\subsection{4.3. \quad Parameter dependence of the scaling relations}
\textcolor{black}{
For each of these scaling relations, we now comment on their dependence and sensitivity to the parameters assumed in our model. In particular, we discuss the sensitivity of these results to the range in values chosen for each of the continuous parameters detailed in Table \ref{Parameter table}: the dark matter halo inner slope, $\alpha$, the fraction of cosmological baryon content, $\eta$, and the intracluster gas inner slope, $\varepsilon$, as well as the dependence on the measurement parameters detailed in our analysis: the detection radius, $r_\mathrm{det}$, and the aperture radius, $R_\mathrm{ap}$, each as was fixed for observational comparison. These parameter dependencies are discussed in each of the headings below, with reference to each of the four predicted scaling relations. }

\textcolor{black}{
In terms of the fixed parameters in Table \ref{Parameter table}: the concentration, $c$, and the dilution, $d$, we offer a brief discussion here as to their impact on our results. The dependence of our predictions on these parameters is best discussed with reference to Figure \ref{Fig - temperature profiles}, tracing the scale-free temperature profiles of our model, with each panel corresponding to the two overdensity conventions: $\Delta=500$ in the top panels, with $c=2.5$ and $d=1$, and $\Delta=200$ in the bottom panels, with $c=5$ and $d=2$. In this figure, the change in profiles between the top panel and the bottom panel is clearly a translation shift of each temperature profile toward smaller halocentric radii, and an amplitude increase for all profiles, except those that exhibit a peak in temperature beyond the meeting point of the profiles, in which case the amplitude is decreased. This quantifiable change in the temperature profiles is mathematically driven by the concentration, which is thus important for the normalisation of each scaling relation with its measurement parameter, as this choice in $c$ sets the radial translation, and will thus impact the calibration of $r_\mathrm{det}$ and $R_\mathrm{ap}$ to the virial radii $r_{200}$ and $r_{500}$ in the emission observables. This general effect can be followed through by comparing the top and bottom panels of Figures \ref{Fig - TX profiles}, \ref{Fig - Tmg profiles}, \ref{Fig - Ysph profiles} and \ref{Fig - Ycyl profiles}, tracing the profiles of each of these observables. }

\textcolor{black}{
To mitigate this potential for introducing a quantitative measurement offset, more accurate modelling could calibrate to specific choices in $c$ on a cluster-by-cluster basis, as informed by the mass-concentration relation \autocite[e.g.][]{bullock2001, duffy2008, ludlow2012, ludlow2014, Correa2015}; however, in these relations $c$ is not a strong function of halo mass in the cluster mass regime, and so any variation in its values is not expected to propagate drastically from our predictions. This avenue was not pursued in our study, as fixing these parameters $c$ and $d$ allowed us to explore the predictions of our model when instead varying the composition and configuration of clusters, whilst maintaining a self-similar approach.
}

\subsubsection{Dependence of the $M_\mathrm{vir} - T_\mathrm{X}$ scaling relation}
\textcolor{black}{
We first comment on the dependence of the $M_\mathrm{vir} - T_\mathrm{X}$ scaling relation before considering the excision of a central core, and so} we refer to the top row of Figure \ref{Fig - weighted X-ray temperature profiles}, which illustrates the dependence of the parameter $\tau_1 \equiv T_\mathrm{X}(<r_\mathrm{det})/T_\mathrm{vir}$ on the continuous structural parameters from Table \ref{Parameter table}. In these panels, the rate of change of $\tau_1$ with respect to the halo inner slope $\alpha$ (the gradient of the curves in these panels) is greatest at the largest, cuspiest value of $\alpha=1.5$ in the permitted range: setting the value of $\tau_\mathrm{1, max}$ in both regimes. This imposes a strong dependence of the $M_\mathrm{vir} - T_\mathrm{X}$ scaling relation on the maximum value of $\alpha$ chosen in the parameter space. Furthermore, the value $\tau_\mathrm{1, min}$ is set by the cuspiest gas inner slope of $\varepsilon=1$ within $\alpha\simeq 0$ halo cores. As $\alpha<0$ halos are unphysical, this minimum value of $\tau_1$ is most sensitive to the maximum value of $\varepsilon$ chosen in the parameter space, implicating the dependence of the $M_\mathrm{vir} - T_\mathrm{X}$ scaling relation on the cuspiest permitted gas profile. For each of these $\tau_1$ curves in Figure \ref{Fig - weighted X-ray temperature profiles}, varying the gas content between $\eta=0.6$ and $\eta=1$ has a negligible impact on the value of $\tau_1$, with this change only barely noticeable for the cuspiest, $\alpha \simeq 1.5$, gas cores, $\varepsilon=0$; thus, there is essentially no dependence of this scaling relation on the gas content of the cluster. 

In terms of the detection radius, $r_\mathrm{det}$, it is clear from the profiles shown in Figure \ref{Fig - TX profiles} that there is almost no dependence of the value of $\tau_1$ on this choice for values $r_\mathrm{det} \gtrsim r_\mathrm{vir}$, except marginal dependence in the case of $\alpha\simeq 0$ halo cores. This independence of the temperature $T_\mathrm{X}$ on the detection radius at large halocentric radii toward $r_\mathrm{det} \simeq r_{200}$ is mathematically expected for NFW-like profiles \autocite[see, e.g.][]{Komatsu&Seljak2001}, which is well-reproduced in the panels of Figure \ref{Fig - TX profiles}.

\textcolor{black}{
When considering the $M_\mathrm{vir} - T_\mathrm{X}$ scaling relation when $T_\mathrm{X}$ is measured with an excised core, these aforementioned parameter dependencies are much more strongly constrained. As seen in Figure \ref{Fig - core-excised temperature profiles}, tracing $\tau_1 \equiv T_\mathrm{X} [r_\mathrm{ex} - r_\mathrm{det}]/T_\mathrm{vir}$ over the same parameter space, it is clear that $\tau_1$ becomes almost insensitive to the gas inner slope, $\varepsilon$, particularly beyond halo cores $\alpha \simeq 0$. The dependence on the maximal value of $\alpha$ remains, as the gradient is still steepest at this boundary; however, this gradient attains a much lower value than previously, and so this dependence is significantly reduced. Moreover, the dependence on the gas content, as in the value of $\eta$, is completely invisible in these curves. This reinforces the importance of core-excision in calibrating the $M_\mathrm{vir} - T_\mathrm{X}$ scaling relation, as sensitivities to both the dark matter and the gas distribution are very strongly reduced. }

\subsubsection{Dependence of the $M_\mathrm{vir} - T_\mathrm{m_g}$ scaling relation}

In the bottom row of Figure \ref{Fig - weighted X-ray temperature profiles}, the dependence of the parameter $\tau_2 \equiv T_\mathrm{m_g}(<r_\mathrm{det})/T_\mathrm{vir}$, and thus the $M_\mathrm{vir} - T_\mathrm{m_g}$ scaling relation, on the halo's structural parameters can be assessed. Unlike for $\tau_1$, there is no consistent trend in which the gradient of these curves is maximised for a particular halo inner slope $\alpha$, despite its value clearly being sensitive to this parameter. Moreover, in these panels it is clear that varying the gas inner slope between $\varepsilon=0$ and $\varepsilon=1$ has minimal impact on the value of $\tau_2$, whilst the effect of varying the gas content between $\eta =0.6$ and $\eta=1$ is completely invisible. As such, the $M_\mathrm{vir} - T_\mathrm{m_g}$ scaling relation is almost completely independent of the gas structure and composition over this parameter space.

Of most sensitivity to this scaling relation is the choice in measurement parameter, the detection radius, $r_\mathrm{det}$. This is illustrated by Figure \ref{Fig - Tmg profiles}, where the $\tau_2$ profiles do not flatten off for detection radii at large halocentric radii, unlike the behaviour of the $\tau_1$ profiles of Figure \ref{Fig - TX profiles}. This imposes a strong sensitivity when calibrating the $M_\mathrm{vir} - T_\mathrm{m_g}$ scaling relation at a fixed detection radius (i.e. when $r_\mathrm{det}=r_{500}$), as, if observational measurements are not precisely measuring this detection radius in a given cluster, the assumed correspondence will be compromised. 

\subsubsection{Dependence of the $M_\mathrm{vir} - Y_\mathrm{sph}$ scaling relation}

To discuss the $M_\mathrm{vir} - Y_\mathrm{sph}$ scaling relation, we refer to Figure \ref{Fig - Ysph profiles}, tracing the SZ parameter $\zeta_1 \equiv Y_\mathrm{sph}(<r_\mathrm{det})/Y_\mathrm{vir}$ over the outlined parameter space. In this figure, varying the gas inner slope between the values $\varepsilon=0$, $\varepsilon=0.5$ and $\varepsilon=1$ across each row entails no clear variation in the profiles traced within each box. In contrast, varying the gas content between $\eta=0.6$ and $\eta=1$ for each of the coloured curves, each individually tracing a fixed halo inner slope, $\alpha$, creates a large interval (shown by the shaded colour regions) that increases in size toward higher detection radii. These intervals are significantly overlapping, such that the values predicted by varying each of $\eta$ and $\alpha$ are strongly degenerate. At detection radii $r_\mathrm{det} \simeq r_\mathrm{vir}$, the intervals spanned by varying $\eta$ dominate the contribution to the bounding region of $\zeta_1$. These trends impose a strong dependence of the $M_\mathrm{vir} - Y_\mathrm{sph}$ scaling relation on the gas content of clusters, but not on the gas profile itself. As such, if values of $\eta$ for galaxy clusters could be more narrowly bounded than $\eta \in [0.6, 1]$, this scaling relation would be significantly more constrained.

In terms of the measurement parameter in this scaling relation, $r_\mathrm{det}$, in each panel of Figure \ref{Fig - Ysph profiles} the values of $\zeta_1$ are a strongly increasing function of increasing $r_\mathrm{det}$. As such, these predictions are strongly sensitive to the chosen value of $r_\mathrm{det}$, implying these relations will be compromised outside of this chosen correspondence. 

\subsubsection{Dependence of the $M_\mathrm{vir} - Y_\mathrm{cyl}$ scaling relation}

Finally, Figure \ref{Fig - Ycyl profiles} allows us to discuss the $M_\mathrm{vir} - Y_\mathrm{cyl}$ scaling relation, by tracing the SZ parameter $\zeta_2 \equiv Y_\mathrm{cyl}(<R_\mathrm{ap})/Y_\mathrm{vir}$ over the outlined parameter space. The same general trends observed for the parameter dependence of $\zeta_1$ are replicated in these panels --- as expected, as both of these observables reflect the same physical dependence of the Compton parameter, only integrated over different volumes. As such, the $M_\mathrm{vir} - Y_\mathrm{cyl}$ scaling relation is strongly dependent on the gas content of clusters, but not the gas profile itself. The values of $\zeta_2$ in Figure \ref{Fig - Ycyl profiles}, traced over a two-dimensional aperture radius, $R_\mathrm{ap}$, are strongly increasing with increasing $R_\mathrm{ap}$; therefore, there is again a strong coupling of this scaling relation to this measurement parameter, with these predictions compromised outside the chosen values of $R_\mathrm{ap}$.

\subsection{4.4. \quad Limitations of the model}

\textcolor{black}{
The results presented in this section, for the scaling relations of galaxy clusters with X-ray and SZ observables, are all dependent upon the assumption of self-similarity. This is built into the derivation of these scaling relations, and incorporated into each of the profiles modelled for the ideal baryonic cluster halos, which are all composed in scale-free, and hence self-similar, form. In the real universe, the scaling of galaxy clusters will deviate from self-similarity, particularly with halo mass, toward the low cluster mass regime, and with high redshift.}

\subsubsection{Low cluster mass limitations}

\textcolor{black}{
When the temperature of the hot ionised gas falls below $\sim 10^{7} \mathrm{K}$, Bremsstrahlung will no longer dominate the cooling mechanism of galaxy clusters \autocite[see, e.g.][]{Sarazin1988}. Below this regime, typically at cluster masses of less than $M_{200} \sim$ a few $ \times 10^{14} \mathrm{M}_\odot$, the cooling function in our definition of $T_\mathrm{X}$, from Equation \eqref{X-ray emission-weighted temperature}, will become an increasingly unreliable model of the gas emission, breaking the assumed self-similarity in this observable. As such, at these low cluster masses, our predictions for the $T_\mathrm{X} - M_\mathrm{vir}$ scaling relation will become unreliable. Without any explicit dependence on the cooling mechanism of the gas or its emission processes, our predictions for the $T_\mathrm{m_g} - M_\mathrm{vir}$ scaling relation, and both of the SZ scaling relations, are not expected to strongly deviate from self-similarity in the low cluster mass regime. This self-similarity is corroborated by observational constraints, over a wide range of cluster masses, whereby slight deviations from self-similarity at the low cluster mass regime can be sufficiently explained in terms of a decrease in the gas fraction, increasing gas clumpiness, or changes in the slope of the gas' pressure profile \autocite[e.g.][]{Ettori2015, Ettori2022}. These differences all reflect changes in the gas' composition or increasing non-gravitational feedback processes in the low cluster mass regime, all of which could be accounted for by adjusting the value of $\eta$, or introducing non-thermal pressure (in modifying the gas' pressure slope), whilst maintaining our general analytic approach. }

\textcolor{black}{
Furthermore, the parameter choices within our model were all strongly informed by observational fits to the intracluster gas density, which represent only the X-ray emitting ionised gas component of galaxy clusters, at temperatures above $\sim$ a few $\times 10^6 \mathrm{K}$. As such, the intracluster gas below such temperatures is not well constrained observationally, and could become significant toward low cluster masses. For example, the gas fraction of galaxy clusters is observed to fall below $\eta = 0.6$, our chosen minimum value, at cluster masses less than $M_{200} \sim $ a few $\times 10^{13} \mathrm{M}_\odot$ (Dev et al. 2024, Submitted). Whether lower temperature ionised gas, below detection capabilities (sometimes known as the `warm' gas component) compensates for these lower gas fractions in low mass clusters is not confidently known. }

\subsubsection{High redshift limitations}

\textcolor{black}{
In our analysis, the form of the cluster's scale-free profiles were all devised in terms of a redshift $z=0$ normalisation, scaled by the cluster's present-day virial parameters, each defined in terms of the present-day critical density, $\rho_\mathrm{crit,0}$. This analysis would need to be corrected for application to high redshift; in particular, re-scaling each profile by redshift-dependent virial parameters, in terms of the critical density $\rho_\mathrm{crit}(z)$ at redshift $z$, which will increase toward earlier times. This will result in a comparatively higher gas density, and a subsequently higher gas temperature and pressure; however, the analytic form of the scale-free profiles will remain unadjusted with this re-scaling. Instead, when maintaining the halo mass conventions $M_{200}$ and $M_{500}$, each cluster at a given halo mass will be able to maintain denser, and thus hotter, gas; this will shift the scaling relations to predict a comparatively smaller halo mass at a given cluster temperature or SZ signal. Beyond this linear adjustment, the parameters in our model were all informed by low-redshift observations, and redshift evolution of some of these parameters --- in particular, the halo concentration \autocite{bullock2001, duffy2008, ludlow2012, ludlow2014} --- is expected. As such, a wider parameter space would need to be physically motivated to encompass high redshift clusters.}

\textcolor{black}{
More limiting for the application of our model, toward higher redshift, is the more dynamic and chaotic state of galaxy clusters at earlier times. In particular, our model predicts the scaling of clusters in virial and hydrostatic equilibrium; at high redshift, both of these equilibria are expected to be more commonly violated, as attributed to higher merger rates \autocite[see, e.g.][]{Genel2009}, in the case of virial equilibrium, and higher gas accretion rates \autocite[see, e.g.][]{Pizzardo2023}, in the case of hydrostatic equilibrium. For these such clusters, our model will not reliably predict their scaling relations, and thus our approach will not be applicable to a larger population of galaxy clusters toward earlier times. }

\section{5. \quad Conclusion}

This study has explored the relationship between a galaxy cluster's dark matter halo mass, the density profile of its dark matter and intracluster gas components, and its intracluster gas emission properties. In particular, we have demonstrated that an analysis of the cluster's emission profiles, in an idealised framework --- referred to as the ideal baryonic cluster halos --- can be used to place constraints on the cluster's scaling relations with these observables. These predictions are thus independent of any numerical models, simulation calibrations or empirical best-fits. 

As a fundamental component of this analysis, we have constructed a competitive tool-kit for modelling the density profile of galaxy clusters. As entirely determined by a minimal set of physically-grounded parameters --- the dark matter halo inner slope, $\alpha$, the halo concentration parameter, $c$, the fraction of cosmological baryon content, $\eta$, the dilution parameter, $d$ and the gas inner slope, $\varepsilon$ --- this analytic model captures the expected structural properties of galaxy clusters, whilst modelling its dark matter and intracluster gas structural profiles in a decoupled and scale-free formalism. We hope this tool-kit can be applied to upcoming X-ray and SZ surveys when modelling the radial profiles of clusters, to better inform these fits in terms of physically-understood parameters. 

As a result of this study, we demonstrated that the halo masses $M_{200}$ and $M_{500}$ can be predicted with an uncertainty of $57.3\%$ and $41.6\%$, respectively, from the emission-weighted temperature observable, $T_\mathrm{X}$, \textcolor{black}{and that these uncertainties reduce by a factor of approximately half, to $31.3\%$ and $17.1\%$, respectively, when a central core is excised from $T_\mathrm{X}$.} Similarly, we predicted these halo masses with a systematically tighter constraint of $25.7\%$ and $7.0\%$, respectively, when recovered from the mean gas mass-weighted temperature observable, $T_\mathrm{m_g}$. Within the parameter space, we demonstrated that the maximum value permitted for the halo inner slope, taken as $\alpha=1.5$, as well as the maximum value permitted for the gas inner slope, taken as $\varepsilon = 1$, impose the strongest parameter dependence for the $M_{200} - T_\mathrm{X}$ and $M_{500} - T_\mathrm{X}$ scaling relations. \textcolor{black}{These parameter dependencies are substantially reduced after excising a central core from the temperature $T_\mathrm{X}$, verifying the importance of this excision in constraining these relations.} In contrast, for the $M_{200} - T_\mathrm{m_g}$ and $M_{500} - T_\mathrm{m_g}$ scaling relations, this sensitivity to the cluster's underlying structure is minimal, producing the resulting stronger constraint on the halo mass. However, for these latter relations, there is a strong dependence on the detection radius, $r_\mathrm{det}$, within which $T_\mathrm{m_g}$ is measured. As the observable $T_\mathrm{X}$ is essentially independent of $r_\mathrm{det}$ when taken at typical values, its scaling relations are correspondingly independent of this measurement. 

Moreover, we predicted the halo masses $M_{200}$ and $M_{500}$ from the cluster's SZ observables: recovered with uncertainties of $31.1\%$ and $17.7\%$, respectively, from the spherically-integrated Compton parameter, $Y_\mathrm{sph}$, and with uncertainties of $25.2\%$ and $22.0\%$, respectively, from the cylindrically-integrated Compton parameter,$Y_\mathrm{cyl}$, such that both SZ observables produce comparable constraints on the halo mass. Within the parameter space, we showed that these scaling relations are insensitive to the form of the gas profile, but instead are driven by the gas content: captured by the fraction of cosmological baryon content $\eta$, with these scaling relations strongly dependent on the minimum $\eta = 0.6$ and maximum $\eta=1$ values chosen in our model. Furthermore, we demonstrated that these scaling relations implicate a strong association with the measurement parameters, either the spherical detection radius, $r_\mathrm{det}$, or the cylindrical aperture radius, $R_\mathrm{ap}$, as both SZ observables are a strong function of increasing integration volume. 

In conclusion, comparing these predictions for the recovery of the halo mass from a cluster's emission observables, the mean gas mass-weighted temperature, $T_\mathrm{m_g}$, and both integrated Compton parameters, $Y_\mathrm{sph}$ and $Y_\mathrm{cyl}$, all provide the strongest constraints, each recovering the halo mass within $\sim 20-30\%$; \textcolor{black}{comparable constraints are also attained for the emission-weighted temperature, $T_\mathrm{X}$, only after excision of the central core.} Furthermore, if the detection radius can be accurately measured , $T_\mathrm{m_g}$ is expected to permit the strongest constraint on the halo mass, as exemplified by an uncertainty of merely $7\%$ when this observable is measured within $r_\mathrm{det} = r_{500}$. Whilst entailing a larger uncertainty, $T_\mathrm{X}$ evades a dependence on accurately knowing the detection radius, which may be preferable observationally when such accuracy is not available. 

Whilst dark matter halos, galaxy clusters and their emission properties remain the focus of increasingly resolved hydrodynamic simulations, we have shown that an analytical framework can be utilised to inform and complement observational constraints on the properties of these complex and dynamic structures. We hope that our predictions will contribute toward improved halo mass estimates, exciting further constraints on the Halo Mass Function and in turn the prospect of untangling the nature of dark matter.

This paper is the second in a series of papers predicting the scaling relations of dark matter halos. In future, we will endeavour to analytically investigate the non-thermal pressure composition of galaxy clusters, which was neglected in this study, to improve this model and contribute toward better understanding the hydrostatic bias.


\vspace{1cm}

 AS acknowledges the support of the Australian Government Research Training Program Fees Offset; the Bruce and Betty Green Postgraduate Research Scholarship; and The University Club of Western Australia Research Travel Scholarship. AS and CP acknowledge the support of the ARC Centre of Excellence for All Sky Astrophysics in 3 Dimensions (ASTRO 3D), through project number CE170100013.

 CB gratefully acknowledges support from the Forrest Research Foundation.

\printbibliography

\end{document}